\documentclass[%
superscriptaddress,
nofootinbib,
amsmath,amssymb,
 aps,
 prd,
floatfix,
]{revtex4-1}
\usepackage{amsmath,latexsym}
\usepackage{bm,hyperref} 
\usepackage{amssymb}
\usepackage{mathtools}
\usepackage{graphicx}
\usepackage{dcolumn}
\usepackage{natbib}
\usepackage[noabbrev]{cleveref}
\begin{document}
\newcommand{\cspace}[2]{{}^{\star}\ensuremath{\mathbb{T}_{#1}\rm#2}}
\newcommand{\tspace}[2]{\ensuremath{\mathbb{T}_{#1}\rm#2}}
\newcommand{\tbundle}[1]{\ensuremath{\mathbb{T}\mathcal{#1}}}
\newcommand{\cbundle}[1]{{}^{\star}\ensuremath{\mathbb{T}\mathcal{#1}}}
\newcommand{\nn}{\nonumber}
\def\a{\alpha}
\def\b{\beta}
\def\g{\gamma}
\def\d{\delta} 
\def\r{\rho}
\def\s{\sigma}
\def\c{\chi}
\newcommand{\p}{\partial}
\newcommand{\lieder}[2]{\mathcal{L}_{#1}{#2}}
\newcommand{\christoffel}[3]{\Gamma^{#1}_{#2#3}}
\newcommand{\at}[2][]{#1|_{#2}}
\newcommand{\inv}[1]{\frac{1}{#1}}
\def\epsUabgd{\epsilon^{\alpha \beta \gamma \delta}}
\def\epsLmnbg{\epsilon_{\mu\nu\beta\gamma}}
\def\epsUmnbg{\epsilon^{\mu\nu\beta\gamma}}
\def\epsLmnab{\epsilon_{\mu\nu\alpha\beta}}
\def\epsUmnab{\epsilon^{\mu\nu\alpha\beta}}

\def\epsUmnrs{\epsilon^{\mu\nu\rho \sigma}}
\def\epsUlnrs{\epsilon^{\lambda \nu\rho \sigma}}
\def\epsUlmrs{\epsilon^{\lambda \mu\rho \sigma}}

\newcommand{\mcA}{{\mathcal{A}}}
\newcommand{\mcB}{{\mathcal{B}}}
\newcommand{\mcC}{{\mathcal{C}}}
\newcommand{\mcD}{{\mathcal{D}}}
\newcommand{\mcE}{{\mathcal{E}}}
\newcommand{\mcF}{{\mathcal{F}}}
\newcommand{\mcG}{{\mathcal{G}}}
\newcommand{\mcH}{{\mathcal{H}}}
\newcommand{\mcI}{{\mathcal{I}}}
\newcommand{\mcJ}{{\mathcal{J}}}
\newcommand{\mcK}{{\mathcal{K}}}
\newcommand{\mcL}{{\mathcal{L}}}
\newcommand{\mcM}{{\mathcal{M}}}
\newcommand{\mcN}{{\mathcal{N}}}
\newcommand{\mcO}{{\mathcal{O}}}
\newcommand{\mcP}{{\mathcal{P}}}
\newcommand{\mcQ}{{\mathcal{Q}}}
\newcommand{\mcR}{{\mathcal{R}}}
\newcommand{\mcS}{{\mathcal{S}}}
\newcommand{\mcT}{{\mathcal{T}}}
\newcommand{\mcU}{{\mathcal{U}}}
\newcommand{\mcV}{{\mathcal{V}}}
\newcommand{\mcW}{{\mathcal{W}}}
\newcommand{\mcX}{{\mathcal{X}}}
\newcommand{\mcY}{{\mathcal{Y}}}
\newcommand{\mcZ}{{\mathcal{Z}}}

\newcommand\maketabularEx{
\begin{tabular}{c|ccc}
0 & 0& 0&0\\
1& 1&0&0\\
1/2&1/4&1/4&0\\
\hline
&1/6&1/6&2/3
\end{tabular}}

\newcommand\maketabularIm{
\begin{tabular}{c|ccc}
$\gamma$ & $\gamma$& 0&0\\
1-$\gamma$& 1-2$\gamma$&$\gamma$&0\\
1/2&1/2-$\gamma$&0&$\gamma$\\
\hline
&1/6&1/6&2/3
\end{tabular}\\\hspace{0.25\textwidth}
$\mathrm{where~} \gamma\equiv1-1/\sqrt{2}$}

\newcommand{\MeV}{\mathop{\rm MeV}}
\newcommand{\fm}{\mathop{\rm fm}}
\newcommand{\GeV}{\mathop{\rm GeV}}
\newcommand{\GeVfmc}{\mathop{\rm GeV/fm^3}}
\newcommand{\TeV}{\mathop{\rm TeV}}
\newcommand{\fmc}{\mathop{\rm fm/c}}
\newcommand{\ten}[1]{10^{#1}}
\newcommand{\diag}{\mathop{\mathrm{diag}}}
\title{Charge diffusion in relativistic resistive second-order dissipative magnetohydrodynamics}

\author{Ashutosh Dash}
\affiliation{Institut f\"ur Theoretische Physik, 
	Johann Wolfgang Goethe--Universit\"at,
	Max-von-Laue-Str.\ 1, D--60438 Frankfurt am Main, Germany}

 \author{Masoud Shokri}
\affiliation{Institut f\"ur Theoretische Physik, 
	Johann Wolfgang Goethe--Universit\"at,
	Max-von-Laue-Str.\ 1, D--60438 Frankfurt am Main, Germany}

\author{Luciano Rezzolla}

\affiliation{Institut f\"ur Theoretische Physik, 
	Johann Wolfgang Goethe--Universit\"at,
	Max-von-Laue-Str.\ 1, D--60438 Frankfurt am Main, Germany}
\affiliation{School of Mathematics, Trinity College, Dublin 2, Ireland}
\affiliation{Frankfurt Institute for Advanced Studies,
  Ruth-Moufang-Str. 1, 60438 Frankfurt am Main, Germany}

\author{Dirk H.\ Rischke}
\affiliation{Institut f\"ur Theoretische Physik, 
	Johann Wolfgang Goethe--Universit\"at,
	Max-von-Laue-Str.\ 1, D--60438 Frankfurt am Main, Germany}
\affiliation{Helmholtz Research Academy Hesse for FAIR, Campus Riedberg,\\
Max-von-Laue-Str.~12, D-60438 Frankfurt am Main, Germany}

\date{\today}
\begin{abstract}
We study charge diffusion in relativistic resistive second-order dissipative magnetohydrodynamics. In this theory, charge diffusion is not simply given by the standard Navier-Stokes form of Ohm's law, but by an evolution equation which ensures causality and stability. This, in turn, leads to transient effects in the charge diffusion current, the nature of which depends on the particular values of the electrical conductivity and the charge-diffusion relaxation time. The ensuing equations of motion are of so-called stiff character, which requires special care when solving them numerically. To this end, we specifically develop an implicit-explicit Runge-Kutta method for solving relativistic resistive second-order dissipative magnetohydrodynamics and subject it to various tests.
We then study the system's evolution in a simplified 1+1-dimensional scenario for a heavy-ion collision, where matter and electromagnetic fields are assumed to be transversely homogeneous, and investigate the cases of an initially non-expanding fluid and a fluid initially expanding according to a Bjorken scaling flow.
In the latter case, the scale invariance is broken by the ensuing self-consistent dynamics of matter and electromagnetic fields. However, the breaking becomes quantitatively important only if the electromagnetic fields are sufficiently strong. The breaking of scale invariance is larger for smaller values of the conductivity. Aspects of entropy production from charge diffusion currents and stability are also discussed.
\end{abstract}

\maketitle

\section{\label{sec:intro}Introduction}

In heavy-ion collisions, the charges of the moving nuclei generate strong electromagnetic fields, the magnitude of which can reach the order of $\sim m_\pi^2 \sim 10^4 \, \MeV^2 \sim 10^{18}$ G \cite{Kharzeev:2007jp, Skokov:2009qp, Voronyuk:2011jd, Bzdak:2011yy, Ou:2011fm, Bloczynski:2012en, Bloczynski:2013mca,Voronyuk:2011jd}. Such fields are among the largest ever observed in the Universe. The strong electromagnetic fields present in heavy-ion collisions, in conjunction with the anomaly of Quantum Chromodynamics (QCD), give rise to anomalous transport phenomena such as the chiral magnetic effect (CME) \cite{Fukushima:2008xe,Kharzeev:2015znc}. The CME has been extensively searched for experimentally in recent years \cite{STAR:2009wot,ALICE:2020siw,Zhao:2019hta,Li:2020dwr,Bzdak:2019pkr}, but so far without conclusive evidence for its existence. The main difficulty is that the lifetime of the magnetic field depends strongly on the electrical conductivity of the produced medium \cite{Tuchin:2013ie, Deng:2012pc, Shokri:2019rsc} and may be too short for the proposed observable signatures of the CME to develop. For other interesting phenomena due to the interplay of electromagnetic fields and nuclear matter, see the recent review \cite{Hattori:2022hyo}.

The generic framework that couples the dynamics of electromagnetic fields with that of a relativistic fluid is called relativistic magnetohydrodynamics (MHD) \cite{Alfven1942,anile1989relativistic}. 
It is well known that ordinary hydrodynamics, i.e., without the coupling to gauge fields, can be considered as an effective theory valid in the low-frequency, large-wavelength limit. The expansion parameter of this theory is the Knudsen number, i.e., the ratio of typical microscopic and macroscopic lengthscales. While the microscopic lengthscale is given by the mean free path of particles or the interparticle separation, the macroscopic lengthscale characterizes the spatio-temporal variations of the hydrodynamical fields \cite{DenicolRischkebook}. At zeroth order in this expansion, the fluid is assumed to instantaneously reach local thermodynamical equilibrium everywhere, which corresponds to ideal (non-dissipative) hydrodynamics. Navier-Stokes theory emerges at first order in Knudsen number. In this theory, dissipative currents are given by the corresponding transport coefficients, such as bulk and shear viscosity, as well as particle diffusion coefficient, multiplied with spatial gradients of the hydrodynamical fields. 
Standard MHD appears as a natural extension of Navier-Stokes theory for a conducting fluid near local thermodynamical equilibrium interacting with electromagnetic fields  \cite{Hernandez:2017mch,Huang:2011dc,Finazzo:2016mhm,Grozdanov:2016tdf}.  

However, first-order Navier-Stokes theory is known to be acausal and unstable \cite{Israel:1976tn} \footnote{Recently it was shown that these acausalities and instabilities can be removed by a suitable matching procedure to the local-equilibrium reference state \cite{Bemfica:2019knx}.}. Transient, or second-order dissipative, theories account for
terms of second order in Knudsen number and avoid such problems by introducing evolution equations for the dissipative currents. These evolution equations have been derived from kinetic theory as underlying microscopic theory using method of moments in Ref.\ \cite{Denicol:2012cn}. In transient hydrodynamical theories, the dissipative currents typically relax on finite timescales to their corresponding Navier–Stokes values \cite{Israel:1979wp,Betz:2009zz,Jaiswal:2013fc}, rendering the system causal and stable if the relaxation times are sufficiently long \cite{Pu:2009fj}. For non-polarizable, non-magnetizable fluids, second-order dissipative hydrodynamics has recently been extended to a theory of resistive second-order dissipative MHD in Refs.\ \cite{Denicol:2018rbw,Denicol:2019iyh}. Relativistic resistive second-order dissipative MHD has been also derived in the relaxation-time approximation in Refs.\ \cite{Panda:2020zhr,Panda:2021pvq}. 

So far, numerical simulations are mostly based on ideal MHD, i.e., the non-resistive (infinite-conductivity) limit \cite{Das:2017qfi,Inghirami:2016iru, Inghirami:2019mkc}. In this paper, we perform the first numerical simulations of resistive second-order dissipative MHD.  We note that, from a numerical perspective, this is a so-called stiff problem, i.e., it involves physical timescales which can be much smaller than the timescale on which the fluid-dynamical variables evolve. Using an explicit time-stepping scheme is not feasible in this case, since one would have to choose a time step which is much smaller than the spatial grid size, which may increase the calculation time to an unacceptably long level. Therefore, one solution is to use an implicit time-stepping scheme. Here, we employ an implicit-explicit Runge-Kutta (IMEX) scheme, which we specifically develop for solving resistive second-order dissipative MHD.
We mention that a similar scheme has been recently employed to solve non-resistive second-order dissipative MHD in the context of astrophysical scenarios \cite{Most:2021rhr}. Another recent development is a code which solves resistive first-order dissipative MHD in 3+1 dimensions \cite{Nakamura:2022wqr}. As the current paper represents the first step in solving
resistive second-order dissipative MHD, we consider a simplified geometry where all fields are homogeneous in two spatial directions, rendering the system effectively 1+1-dimensional. We also neglect the effects of bulk and shear viscosity and exclusively focus on the evolution of the charge diffusion current. One motivation for this work comes from recent results of transport simulations \cite{Wang:2021oqq,Yan:2021zjc,Grayson:2022asf}, which reveal that the charge diffusion current needs some time to reach the value given by the standard Navier-Stokes form of Ohm’s law. Our goal is to investigate such transient dynamics via resistive second-order dissipative MHD calculations.

This paper is organized as follows. In Sec.\ \ref{sec:HydroEqn} we provide a short review of resistive second-order dissipative MHD for non-polarizable, non-magnetizable fluids, neglecting the effects of bulk viscous pressure and shear-stress tensor. 
In Sec.\ \ref{sec:NumProcedures} we discuss the principles of our numerical implementation of resistive second-order dissipatve MHD. Then, in Sec.\ \ref{sec:HIC} we discuss our simplified set-up of a relativistic heavy-ion collision and present the results of our numerical simulations.
 Section \ref{sec:ConcandOut} concludes this work with a summary of our results and an outlook.  The Appendix contains further details of our numerical approach as well as the study of various test cases.
 
We use natural Heaviside-Lorentz units in which $\hbar=c=k_B = \epsilon_0 = \mu_0 =1$. The convention for the metric tensor is ``mostly minus'', i.e., $ g_{\mu \nu} = \diag(1,-1,-1,-1)$.
The comoving derivative of a quantity $A$
is denoted as $\dot{A} \coloneqq u^\mu \partial_\mu A$, where $u^\mu$ is the fluid four-velocity. The projector onto the three-space orthogonal to the fluid velocity is defined as $\Delta^{\mu\nu} \coloneqq g^{\mu\nu}-u^\mu u^\nu$. The projection of a vector $A^\mu$ orthogonal to the fluid velocity is denoted as $A^{\langle\mu\rangle} \coloneqq \Delta^{\mu\nu}A_\nu$.
The covariant spatial gradient is denoted as $\nabla^\mu \coloneqq \Delta^{\mu \nu} \partial_\nu$. The symmetric, traceless projector of rank four is $\Delta^{\mu\nu}_{\alpha\beta} \coloneqq \tfrac{1}{2}\left(\Delta^{\mu}_{\alpha}\Delta^{\nu}_{\beta}+\Delta^{\mu}_{\beta}\Delta^{\nu}_{\alpha}\right)-\tfrac{1}{3}\Delta^{\mu\nu}\Delta_{\alpha\beta}$, the application of which onto a rank-2 tensor $A^{\mu\nu}$ is denoted by $A^{\langle\mu\nu\rangle} \coloneqq \Delta^{\mu\nu}_{\alpha\beta}A^{\alpha\beta}$.

\section{Equations of motion} \label{sec:HydroEqn}

The evolution of a relativistic fluid coupled to electromagnetic fields is effectively described by the conservation laws of energy-momentum and conserved charges, Maxwell's equations, and, in standard resistive MHD, by constitutive relations for the dissipative currents. For instance, the diffusive part of the charge current is given by the Navier-Stokes form of Ohm's law, which encodes the instantaneous response of the fluid's diffusion current to an electric field proportional to the electrical conductivity. The latter depends on the underlying microscopic properties of the fluid. 
In resistive second-order dissipative MHD, the constitutive relations are replaced by evolution equations for the dissipative currents, and Ohm's law assumes a more complex form.
In this work, which constitutes the first attempt at solving resistive second-order dissipative MHD numerically in a semi-realistic set-up, we neglect the evolution of bulk viscous pressure and shear-stress tensor and focus solely on the charge diffusion current.
In this section, we describe  in more detail the three sets of equations, namely conservation laws, Maxwell's equations, and the equation for the charge diffusion current.

\subsection{Conservation laws}

The conservation laws of energy-momentum and charge read 
\begin{align}\label{eq:partialTmunu}
\partial_{\mu}T^{\mu\nu}& =0\;, \\
\label{eq:partialrho}
\partial_{\mu}J_f^\mu& =0\;,
\end{align}
where $T^{\mu \nu}$ is the total energy-momentum tensor of the system and $J_f^\mu$ is the charge four-current of the fluid.
The energy-momentum tensor is given by
\begin{equation}\label{eq:Tmunu}
    T^{\mu\nu} =  T^{\mu\nu}_f+T^{\mu\nu}_{em}\;.
\end{equation}
It consists of a fluid part $T^{\mu \nu}_f$ and an electromagnetic part $T^{\mu \nu}_{em}$, which for non-polarizable, non-magnetizable fluids reads~\cite{Misner:1973prb}
\begin{equation}
 T^{\mu\nu}_{em} = -F^{\mu\lambda}F^{\nu}_\lambda+\frac{1}{4}g^{\mu\nu}F^{\alpha\beta}F_{\alpha\beta}\;.
\end{equation}
Here,
\begin{equation}
    F^{\mu\nu}= \mathcal{E}^\mu u^\nu-\mathcal{E}^\nu u^\mu+\epsilon^{\mu\nu\alpha\beta}u_\alpha \mathcal{B}_\beta
\end{equation}
is the Faraday tensor decomposed in terms of the fluid four-velocity $u^\mu$, as well as the electric and magnetic field four-vectors $\mathcal{E}^\mu\equiv F^{\mu\nu}u_\nu$ and $\mathcal{B}^\mu\equiv\frac{1}{2}\epsilon^{\mu\nu\alpha\beta}F_{\alpha\beta}u_\nu$ in the comoving frame, respectively~\cite{PhysRevD.3.2941}.
Neglecting contributions from bulk viscous pressure and shear-stress tensor,
the energy-momentum tensor of the fluid reads in the Landau frame
\begin{equation} \label{eq:T_fmunu}
  T^{\mu\nu}_{f}\equiv wu^\mu u^\nu-Pg^{\mu\nu}\;,
\end{equation}
where $w\equiv\varepsilon+P$ is the enthalpy density, with $\varepsilon$ being the energy density and $P$ the pressure, respectively. It should be noted that, in the presence of an external charge current $J^\mu_{ext}$, the divergence of $T^{\mu \nu}$ is given by
\begin{equation}
\partial_{\mu}T^{\mu\nu}=-F^{\nu}_{\hspace{0.1cm}\lambda} J^\lambda_{ext}\;,
\end{equation}
because external currents induce electromagnetic fields and thus feed energy and momentum into the system. 

For the sake of simplicity we assume that the fluid consists of particles (and antiparticles) with a single charge $q$. The charge current of the fluid is then given by
\begin{equation}{\label{eq:Chconsrveqn}}
	 J^{\mu}_f\equiv q \left(n u^\mu+V^{\mu}\right)\;,
	\end{equation}
where $n$ is the net particle density in the local rest frame of the fluid and $qV^{\mu}\equiv \Delta^{\mu}_{\nu}J_f^\nu$ is the charge diffusion current.
We note that the pressure in Eq.\ (\ref{eq:T_fmunu}) is not an independent variable, but given
by an equation of state (EOS) of the
form $P(\varepsilon,n)$.

\subsection{Maxwell's equations}

The evolution of the electric and magnetic fields is given by Maxwell’s equations,
		\begin{align}
		{\label{eq:maxwelleqn1}}
		\partial_{\mu}F^{\mu\nu} &= J^{\nu}\;, \\{\label{eq:maxwelleqn2}}
		\epsilon^{\mu\nu\alpha\beta}\partial_{\mu}F_{\alpha\beta} &= 0\;,
	\end{align}
where the total charge four-current $J^{\mu}\equiv J^{\mu}_f+J^{\mu}_{ext}$ serves as the source for the electromagnetic fields.

\subsection{Charge diffusion current}

Ohm's law, in its simplest covariant 
Navier-Stokes-type form, reads \cite{Komissarov:2007wk, Palenzuela:2008sf, Takamoto:2011ng} 
\begin{equation}{\label{eq:Ohmslawacausal}}
    qV^\mu=q\kappa\nabla^{\mu} \alpha+\sigma \mathcal{E}^\mu\;.
\end{equation}
Here, $\sigma$ is the electrical conductivity, $\kappa$ is the particle diffusion coefficient, and $\alpha \coloneqq \mu/T$ is the ratio of chemical potential $\mu$ to temperature $T$. The electrical conductivity and charge diffusion coefficient satisfy the  Wiedemann–Franz law, $\sigma=q^2\kappa/T$. The well-known ideal-MHD limit, i.e, when $\sigma\rightarrow\infty$, can be obtained from Eq.\ \eqref{eq:Ohmslawacausal} by retaining only the second term on the right-hand side and demanding $\mathcal{E^\mu}=0$ to have a finite $J^\mu_f$. 

As discussed in the Introduction, the Navier-Stokes form  (\ref{eq:Ohmslawacausal}) of Ohm's law is problematic in a relativistic context, because it allows signals to propagate with infinite speed, violating causality \cite{forster1975hydrodynamic,Aziz:2004qu}. Furthermore, a modification of this standard form of Ohm's law is required on the grounds that, even when an electric field is already present, the build-up of the corresponding charge diffusion current needs a finite time to reach the form given by Eq.\ \eqref{eq:Ohmslawacausal}, as also recently found in transport simulations  \cite{Wang:2021oqq,Yan:2021zjc,Grayson:2022asf}. In this work, we use the equation of motion of the charge diffusion current as derived in resistive second-order dissipative magnetohydrodynamics  \cite{Denicol:2019iyh,Panda:2021pvq}.
In its most simple form it reads 
\begin{equation}{\label{eq:Ohmslaw}}
     \tau_V q \dot{V}^{\langle\mu\rangle}
     + q V^{\mu}=q \kappa\nabla^{\mu} \alpha +\sigma \mathcal{E}^\mu\;,
\end{equation}
where $\tau_V$ is the charge-diffusion relaxation time. Note that the charge diffusion equation \eqref{eq:Ohmslaw} represents the simplest possible way, in the spirit of the Maxwell-Cattaneo construction, to incorporate a causal time lag for the response of the charge diffusion current to the dissipative forces on the right-hand side. Possible generalizations include additional terms as given in Eq.\ (25) of Ref.\ \cite{Denicol:2019iyh}, see also Ref.\ \cite{Panda:2021pvq}. For
vanishing bulk viscous pressure and shear-stress tensor, and for fluids which only move in one spatial dimension, most of these terms are zero. The other,  non-vanishing ones are not necessarily small, but are omitted here to keep the discussion as simple as possible.
Equations (\ref{eq:partialTmunu}), (\ref{eq:partialrho}),
(\ref{eq:maxwelleqn1}), (\ref{eq:maxwelleqn2}), and
(\ref{eq:Ohmslaw}), together with an EOS for the fluid and expressions for the charge diffusion coefficient, conductivity, and relaxation time, completely describe the system under consideration provided consistent initial and boundary data are given.

\subsection{Qualitative features of the resistive MHD description}

At this point, it is useful to discuss some properties of the coupled system
of equations of motion of relativistic resistive second-order dissipative MHD.
To this end, for the sake of simplicity we neglect particle-density gradients, such that
$\nabla^\mu \alpha \equiv 0$, and consider the rest frame of the fluid, in which case Ohm's law \eqref{eq:Ohmslaw} becomes
\begin{equation}
   \tau_V q \dot{V}^{i}=\sigma E^i-qV^{i}\;,
\end{equation}
where $i = 1, 2,$ or 3, and $E^i = F^{i0}$
is the electric field.

Furthermore assuming that the conductivity and relaxation time are constant in space-time, we take the time derivative of the above equation and then use Ampere's law to obtain the equation of
motion of a damped, driven harmonic oscillator,
\begin{equation}\label{eq:DampedOscii}
    \ddot{V}^{i}+2\,\omega_0\, \zeta_d\, \dot{V}^{i}+\omega_0^2 V^i=\frac{\omega_0^2}{q}\, \epsilon^{ijk}\partial_j B_k\;.
\end{equation}
where $\omega_0 \coloneqq \sqrt{\sigma/\tau_V}$, $\zeta_d \coloneqq 1/(2\sqrt{\sigma\tau_V})$,
and $B^i = - \frac{1}{2} \epsilon^{ijk} F_{jk}$.

The value of the damping ratio $\zeta_d$  determines the qualitative behavior of the system. (i) If $\zeta_d>1$, the charge diffusion current is overdamped and exponentially decays without oscillations. (ii) The case $\zeta_d=1$ corresponds to the critical aperiodic-limit case, where the charge diffusion current returns to a steady state as quickly as possible without oscillating. Finally, (iii) if $\zeta_d<1$, the charge diffusion current is underdamped and oscillates with an amplitude which gradually decreases to zero. 

For applications in heavy-ion collisions, let us assume that $\sigma \sim 0.02\, T$, in accordance with lattice-QCD results \cite{Aarts:2020dda}. Then, for case (ii), the relaxation time must be $\tau_V \sim 10/T$. Oscillations only happen if the relaxation time is larger than this value. For $T \sim 200 \MeV$, $\tau_V \sim 10 \fm$, which is comparable to the lifetime of the fireball. For such a large relaxation time, the hydrodynamical description (as considered as an expansion in powers of Knudsen number) breaks down, since then the microscopic scale $\tau_V$ is no longer much smaller than the macroscopic scale, in this case the system's lifetime.

In Sec.~\ref{sec:NumProcedures} and App.\ \ref{sec:IMEX}, we discuss in detail how to numerically solve the system of Eqs.\ (\ref{eq:partialTmunu}), (\ref{eq:partialrho}),
(\ref{eq:maxwelleqn1}), (\ref{eq:maxwelleqn2}), and
(\ref{eq:Ohmslaw}). We subject our numerical procedure to various tests, which are discussed in App.~\ref{App:TestCases}. Applications to heavy-ion collisions are studied in
Sec.\ \ref{sec:HIC}.

\section{Numerical procedure}\label{sec:NumProcedures}
In this section, we present the numerical method to solve the equations of motion of relativistic resistive second-order dissipative MHD. As already advertised, this poses a stiff problem, which requres the use of a dedicated method to obtain a solution which is correct from a mathematical, and therefore also physical, point of view. In the following subsections, we first formulate the equations of motion in a fixed frame, the lab frame, and then discuss hyperbolic equations with stiff terms in general and then, more specifically, in the application to relativistic resistive second-order dissipative MHD.

\subsection{Equations of motion in a fixed frame} 

In order to numerically solve the system of equations of motion of  relativistic resistive second-order dissipative MHD one chooses a frame, in the following called lab frame.
In this frame, the equations of motion (\ref{eq:partialTmunu}), (\ref{eq:partialrho}), (\ref{eq:maxwelleqn1}), (\ref{eq:maxwelleqn2}), and (\ref{eq:Ohmslaw}) can be cast into the following conservative form,
\begin{equation}\label{Eq:systemofeqn}
 \partial_t \boldsymbol{U}+\partial_j \boldsymbol{F}^j(\boldsymbol{U})=\boldsymbol{{S}}(\boldsymbol{U})\;,
\end{equation} 
where $\boldsymbol{U}$ represents the vector of conserved quantities and $\boldsymbol{F}^j$ the vector of fluxes,
\begin{equation}
\boldsymbol{U}=\begin{pmatrix}
  e \\ M^i \\ N_f\\ V^i \\B^i \\E^i
\end{pmatrix}\; , \qquad \boldsymbol{F}^j(\boldsymbol{U})=
\begin{pmatrix}
 F^j_e \\F^{ij}_M \\ v^j N_f\\v^j V^i\\ \epsilon^{ijk}E_k\\-\epsilon^{ijk}B_k
\end{pmatrix}\;,
\end{equation}
and the vector of sources reads
\begin{equation}
\boldsymbol{S}(\boldsymbol{U})=
\begin{pmatrix}
0\\0\\-\partial_i(-v^iV^0+V^i)\\ \left(\sigma\mathcal{E}^i+ q\kappa \nabla^{i}\alpha-qV^{i}\right)/(q\tau_V\gamma)+V^i\partial_j v^j-u^i V^\nu \dot{u}_\nu\\0\\-J_f^i
\end{pmatrix}\;,
\end{equation}
along with constraint equations
\begin{align}\label{eq:constE}
\partial_i E^i&=J^0_f\;,\\\label{eq:constB}
\partial_i B^i&=0\;. 
\end{align}
In the lab frame, the fluid four-velocity is $u^\mu=\gamma(1,\boldsymbol{v})$, where $\gamma=(1-v^2)^{-1/2}$, and the quantities $e$, $M^i$, and $N_f$ appearing in the vector $\boldsymbol{U}$ are given as
\begin{align}
e&= \frac{1}{2}\left(E^2+B^2\right)+\gamma^2 w-P\;,\\
M^i&= \gamma^2 wv^i+\epsilon^{ijk}E_jB_k\;,\\
N_f&= q(n \gamma+V^0)\;.
\end{align}
Note that in the above equations $E^i$ and $B^i$ are the electromagnetic fields as measured in the lab frame whereas $\mathcal{E}^\mu$ and $\mathcal{B}^\mu$ are the electromagnetic fields as measured in the comoving frame of the fluid, respectively. Similarly, the quantities
$F_e^i$, $F_M^{ij}$ appearing in the
flux vector $\boldsymbol{F}^j(\boldsymbol{U})$ 
read
\begin{align}
 F^{i}_e&= M^i\;, \\
 F^{ij}_M&=M^iv^j-Pg^{ij}-E^iE^j-B^iB^j-\frac{1}{2}
 \left(E^2+B^2\right)g^{ij}\;.
\end{align}

\subsection{Hyperbolic equations with stiff terms}

While the equations of motion of ideal MHD can be numerically solved very efficiently, the equations of motion of relativistic resistive second-order dissipative MHD pose considerable difficulties for a numerical solution, in particular when the conductivity in the plasma is large. In regions with high conductivity, the system will evolve on timescales which are very different from those in regions of low conductivity.  Mathematically, we have to deal with
a system of hyperbolic equations with stiff relaxation terms, which requires special care to capture the dynamics in a stable and accurate manner.

A prototype hyperbolic equation with a stiff source term $\boldsymbol{R}(\boldsymbol{U})/\epsilon$ and a non-stiff source term $\boldsymbol{T}(\boldsymbol{U})$ is given by
 \begin{equation}{\label{eq:prototype}}
   \partial_t \boldsymbol{U} + \partial_j \boldsymbol{F}^j(\boldsymbol{U})=\frac{1}{\epsilon} \boldsymbol{R}(\boldsymbol{U})
   + \boldsymbol{T}(\boldsymbol{U})\;,
 \end{equation}
where $\epsilon>0$ is a relaxation time (which is not necessarily identical with $\tau_V$).

For example, let us consider the Navier-Stokes form \eqref{eq:Ohmslawacausal} of Ohm's law, which can be explicitly written as
\begin{equation}
   J^i_f=\sigma\gamma(E^i+\epsilon^{ijk}v_jB_k-v^i E^jv_j)+qn v^i\;.
\end{equation}
In the fluid rest frame, this reduces to the usual form $J_f^i=\sigma E^i$. In this case, there is no equation of motion
 for the charge diffusion current $qV^\mu$, and the system  \eqref{Eq:systemofeqn} of equation of motion has the reduced set of conserved variables $\boldsymbol{U}=(e,M^i,B^i,E^i)$ and flux variables  $\boldsymbol{F}^j=(F^j_e,F^{ij}_M, \epsilon^{ijk}E_k,-\epsilon^{ijk}B_k)$, with sources $\boldsymbol{S}=(0,0,0,-J^i_f)$. Hence, a comparison to the prototype equation \eqref{eq:prototype} reveals that the relaxation time $\epsilon$ can be identified with the resistivity $1/\sigma$.

In the limit $\epsilon \rightarrow \infty$,
the stiff source term in Eq.\ (\ref{eq:prototype}) vanishes.
In this case, the system is equivalent to a standard hyperbolic equation. A typical finite-volume method to numerically solve such a system requires a bound for the speed $c_h$ with which perturbations propagate in the fluid. In standard fluid dynamics, $c_h$ is given by the speed of sound waves propagating relative to the fluid flow, e.g., for
one-dimensional fluid flow, $v^x=v$, $v^y=v^z=0$ \cite{2013rehy.book.....R},
\begin{equation}\label{eq:spectralradius}
c_h=\mathrm{max}\left(\frac{v+c_s}{1+vc_s},\frac{v-c_s}{1-vc_s}\right)\;,
\end{equation}
where $c_s^2 \coloneqq \partial P / \partial \varepsilon |_{s/n}$ (with $s$ being the entropy density of the fluid) is the speed of sound. The speed $c_h$, together with the hydrodynamic lengthscale $L$ defined by the spatio-temporal variations of the hydrodynamic variables, defines a characteristic timescale $\tau_h=L/c_h$ of the hyperbolic part. 

In the opposite limit $\epsilon\rightarrow 0$, corresponding to infinite conductivity, the system is stiff, since the timescale $\epsilon$ of the stiff source term $\boldsymbol{R}(\boldsymbol{U})$ is much smaller than $\tau_h$. In this limit, the stability of a numerical solution scheme using explicit time-stepping can only be achieved choosing the timestep $\Delta t\leq \epsilon$. For small $\epsilon$, such a requirement can considerably prolong calculation times and thus renders an explicit integration scheme impractical.

Therefore, to solve a stiff system of hyperbolic equations, one has to employ a different numerical approach.
Typically, one uses either one of the following methods:
\begin{enumerate}
 \item The Strang-Splitting method, \cite{jahnke2000error} which provides second-order accuracy if each step is second-order accurate. However, higher-order accuracy is difficult to obtain even in non-stiff problems with this kind of method.
 \item  Implicit-explicit Runge-Kutta (IMEX), \cite{Pareschi:2005:IER} which represents an effective solution to the problem when a hierarchy of vastly different timescales is involved and one does not want to sacrifice numerical accuracy. The main drawback is the difficulty of implementation.
\end{enumerate}

We note that the use of the IMEX method in this work has been crucial to obtain the mathematically correct solution in the regimes of large conductivity, where the stiff character of the equations is most evident. Alternative numerical methods, such as the Strang-splitting method, have also been tested but can be shown to be inadequate under the physical conditions explored here. A more detailed discussion of the numerical implementation of the IMEX method for relativistic resistive second-order dissipative MHD is presented in App.~\ref{sec:IMEX}, while
various test cases, including a comparison with the Strang-Splitting method, are presented in App.~\ref{App:TestCases}.

 \section{Simplified set-up for heavy-ion collisions}\label{sec:HIC}
 
     \begin{figure}
     \centering
    \includegraphics[width=1.0\textwidth]{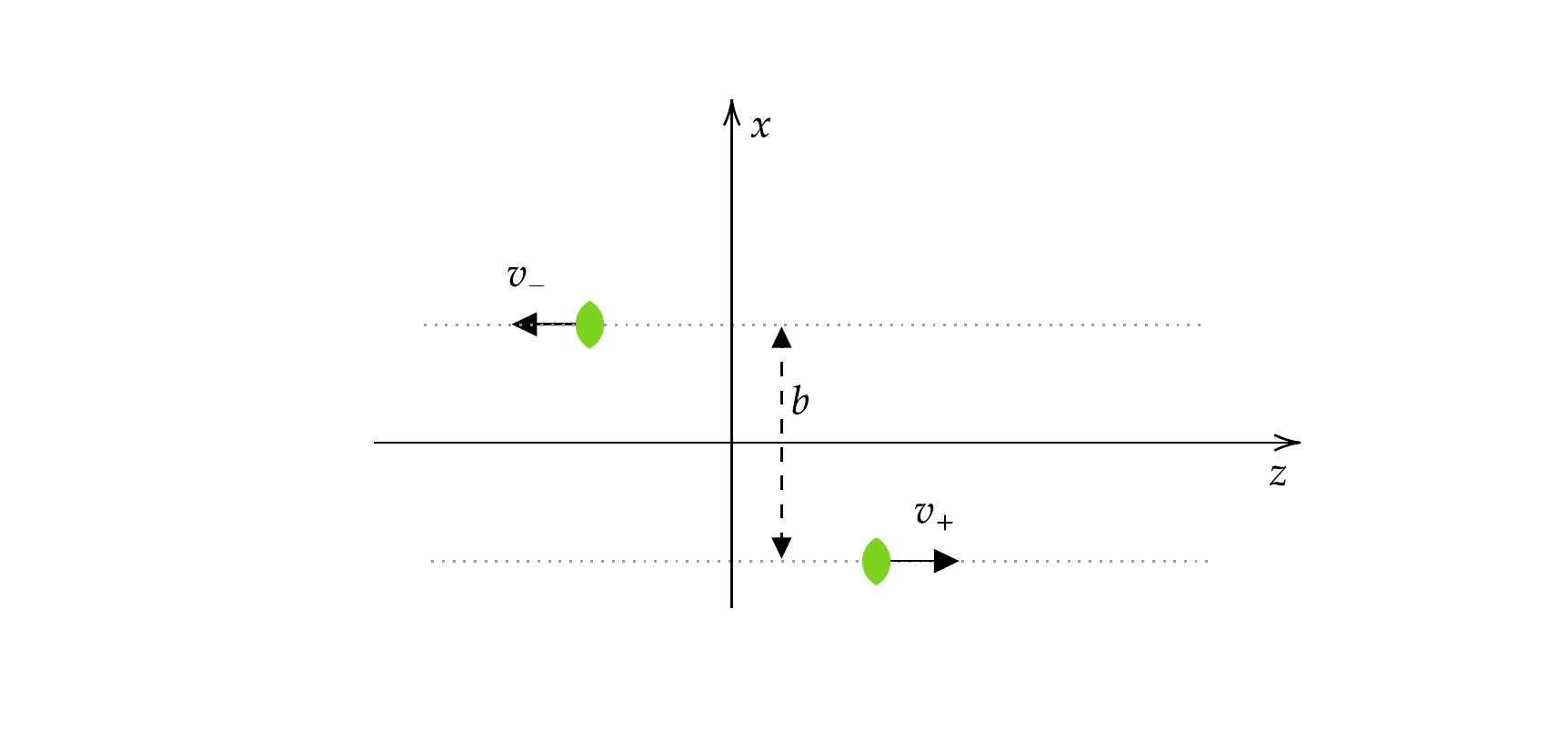}
    \caption{Collision geometry of two Lorentz-contracted nuclei moving with velocity $v_\pm$ in the $\pm z$ direction at impact parameter $b$ along the $x-$axis.}
    \label{fig:geometry}
\end{figure}

 In this section, we apply the reduced MHD approach discussed in Sec.\ \ref{sec:HydroEqn} to a simplified set-up for heavy-ion collisions, where the system is homogeneous in the directions transverse to the beam axis, while the fluid only moves in longitudinal (i.e., beam) direction. 

At relativistic energies, the momenta of the colliding nuclei are so large that the nuclear stopping power is not sufficient to significantly decelerate them. Therefore, we assume that the electromagnetic field is created by  protons moving with rapidity $\pm Y_{\mathrm{bm}}$, where $Y_{\mathrm{bm}}= \mathrm{Artanh} \sqrt{1-4m_N^2/s_{NN}}$ is the beam rapidity in the center-of-momentum (C.M.) frame~\cite{Liu:2014ixa} and
$\pm$ corresponds to the protons moving in $\pm z$-direction.
Here, $m_N$ is the mass of the nucleon and $\sqrt{s_{NN}}$ is the C.M.\ energy per nucleon pair. We do not distinguish between spectator and participant protons, i.e., assume that the electromagnetic field of a nucleus is created by a 
charge of magnitude $Z$. For the sake of simplicity we also assume that charge to be point-like. Then, the electromagnetic four-potential in the Lorenz gauge (see Ref.\ \cite{schwinger2019classical}) can be constructed by boosting the electrostatic potential of a charge at rest to the C.M.\ frame,
\begin{equation}
A_{\pm}^\mu=\left(\frac{Z\alpha_{\mathrm{EM}}\gamma}{r_\pm},0,0,v_\pm\frac{Z\alpha_{\mathrm{EM}}\gamma}{r_\pm}\right)\;,
\end{equation}
where $\alpha_{\mathrm{EM}}$ is the fine-structure constant, $ v_\pm=\pm\tanh Y_{\mathrm{bm}}$ are the velocities of the nuclei, and $\gamma \coloneqq (1-v_{\pm}^2)^{-1/2}$.
At time $t$, the distance between the observer
at $(x,y,z)$ and the charge moving
with velocity $v_\pm$ is $r_\pm (x,y,z,t) \coloneqq \sqrt{(x\pm b/2)^2+y^2+\gamma^2(z-v_\pm t)^2}$ and points from the center of the nucleus at $(\pm b/2, 0, v_{\pm}t )$ to the observer, see Fig.\ \ref{fig:geometry}. 

We assume that the system is homogeneous in the transverse plane, hence consider only the electromagnetic field near $\boldsymbol{x}_{\bot} \coloneqq (x,y) =0$. The non-zero components of the Faraday tensor  $F^{\mu\nu}=\partial^\mu A^\nu-\partial^\nu A^\mu$, where $A^\mu=A_{+}^\mu+A_{-}^\mu$, at
$\boldsymbol{x}_\bot =0$ are then given by
\begin{align}\label{eq:Intialemfields1}
    B^y(0,0,z,t)&=\frac{b}{2}\, Z\alpha_{\mathrm{EM}} \left(\frac{1}{r_{0,+}^3}
    +\frac{1}{r_{0,-}^3}\right)\, \mathrm{sinh} Y_{\mathrm{bm}}\;,\\\label{eq:Intialemfields2}
    E^x(0,0,z,t)&=\frac{b}{2}\, Z\alpha_{\mathrm{EM}} \left(\frac{1}{r_{0,+}^3}
    -\frac{1}{r_{0,-}^3}\right)\, \mathrm{cosh} Y_{\mathrm{bm}}\;,
\end{align}
where $r_{0,\pm} \coloneqq r_\pm (0,0,z,t)$.

In the following, we will consider two cases. The first one is an initially non-expanding fluid, i.e., the fluid is at rest with a uniform energy-density profile in $z$-direction. This is done in order to compare with previous studies \cite{Tuchin:2013ie,Gursoy:2014aka}, where the authors solve Maxwell's equations in conducting media. The second case is a fluid which is initially expanding in $z$-direction according to the boost-invariant Bjorken-flow scenario, i.e., with velocity $v^z=z/t$.

\subsection{Initially non-expanding fluid}
\label{sec:IIIA}
We consider Au-Au collisions at $\sqrt{s_{NN}}=200\GeV$ with impact parameter $b=10\fm$. The initial pressure is $P=18.33\GeVfmc$ and the initial electromagnetic fields are calculated using Eqs.\ (\ref{eq:Intialemfields1}) and (\ref{eq:Intialemfields2}) at time $t=10^{-3}\fm$. All other fields are initially set to zero, including the initial net charge density so that the constraint Eq.\ \eqref{eq:constE} is identically satisfied. We use the EOS $P=\varepsilon/3$, i.e., $c_s^2=1/3$.  However, in principle other EOSs can also be taken. The electrical conductivity $\sigma$ and relaxation time $\tau_V$ are kept as free parameters. Varying them we will study the cases where
$\zeta_d < 1$ or $\zeta_d >1$.
   \begin{figure}
    \centering
    \includegraphics[width=0.48\textwidth]{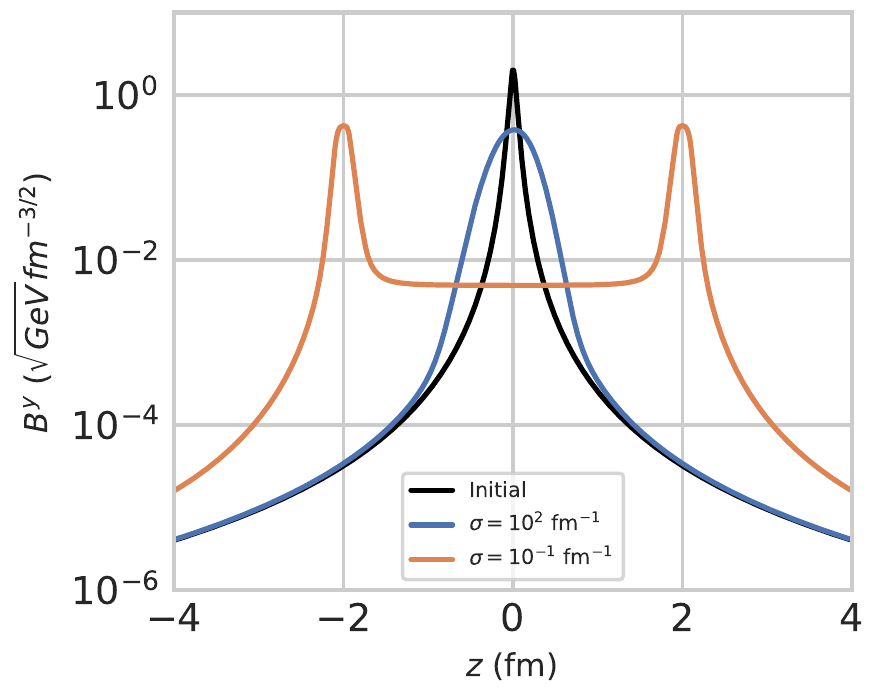}
    \includegraphics[width=0.48\textwidth]{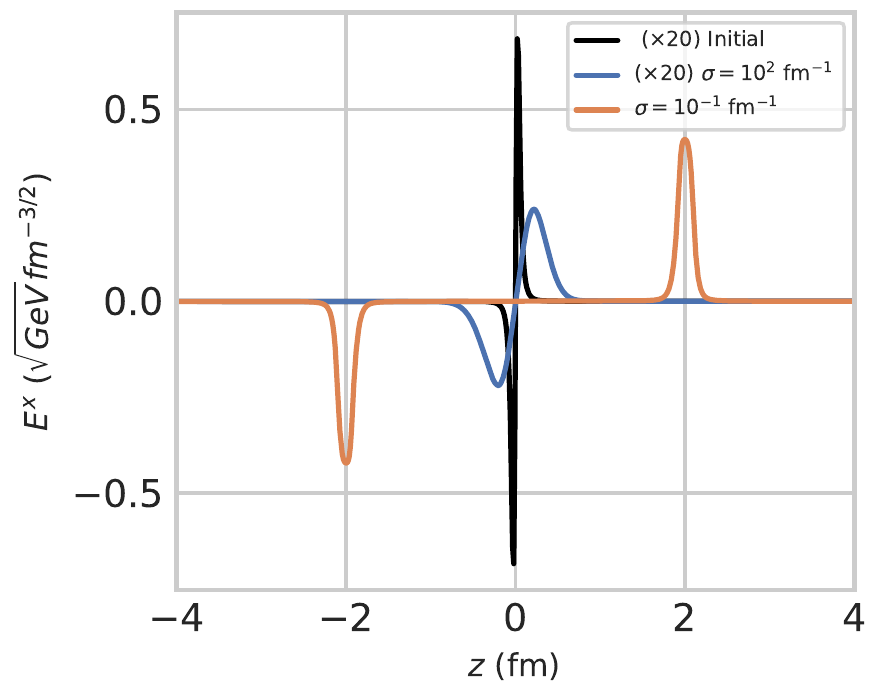}
    \caption{Left panel: The magnetic-field component $B^y$ as a function of $z$ for the initially non-expanding case. The black solid line is the initial magnetic field, while the other two lines are for two different values of the conductivity at time $t=2$ fm. The relaxation time is fixed to $\tau_V=10^{-2}$ fm. Right panel: Same as left panel but for the electric-field component $E^x$.}
    \label{fig:NonexpandingEmfields}
\end{figure}
   \begin{figure}
    \centering
    \includegraphics[width=0.48\textwidth]{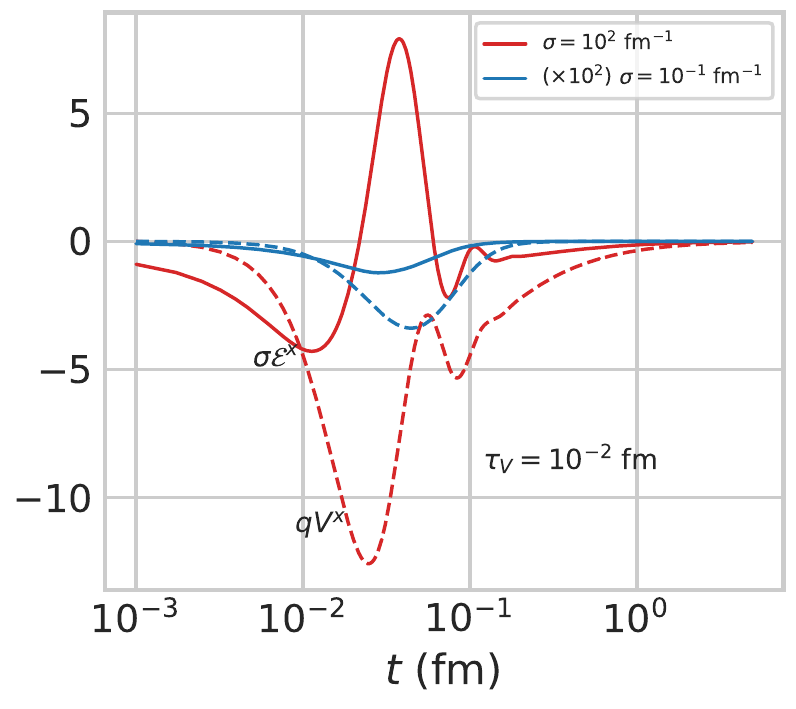}
    \includegraphics[width=0.487\textwidth]{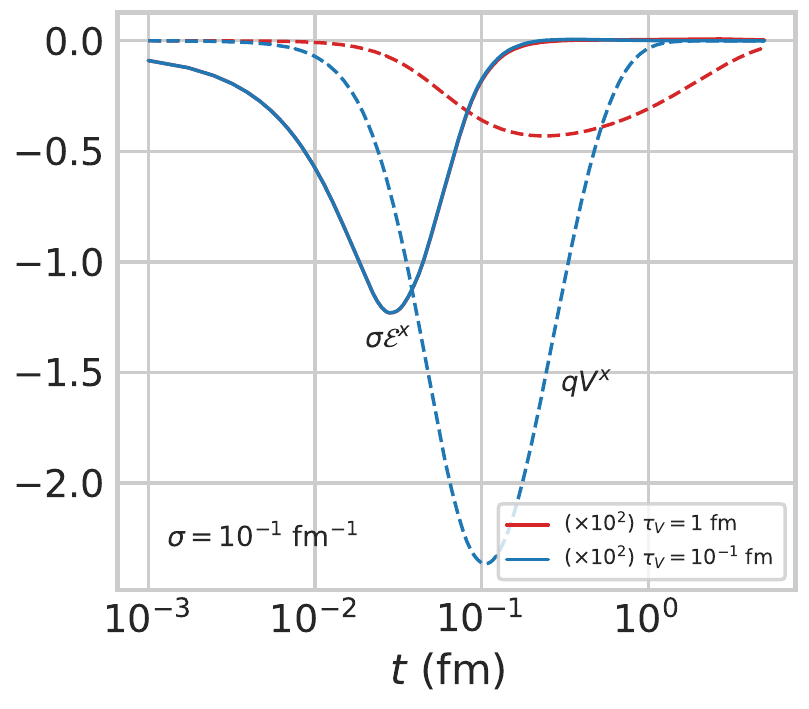}
    \caption{Left panel: Time evolution of the charge diffusion current $qV^x$ (dashed line) and $\sigma \mathcal{E}^x$ (solid line) at $z=-0.006$ fm for two different values of $\sigma$, but for the same $\tau_V=10^{-2}$ fm. Right panel: Same as left panel but for different values of $\tau_V$ for the same $\sigma=10^{-1}$ fm$^{-1}$.}
    \label{fig:JBysigmaE}
\end{figure}

Figure \ref{fig:NonexpandingEmfields} shows the $z$-dependence of the electromagnetic fields at time $t=2$ fm for different values of $\sigma$ and fixed $\tau_V=10^{-2}$ fm. From the left panel of Fig.\ \ref{fig:NonexpandingEmfields} one observes that the magnetic field is even under $z \rightarrow - z$. For large $\sigma$, the field deviates only slightly from the initial profile of the magnetic field. This is due to the frozen-flux theorem, which tells us that the magnetic field is confined inside the matter, which is, at least initially, not expanding.
For small $\sigma$, the magnetic-field evolution is closer to that in vacuum, i.e., it is peaked at the positions of the moving charges, with small diffusive tails which become narrower as $\sigma \rightarrow 0$. The right panel of Fig.\ \ref{fig:NonexpandingEmfields} shows that the electric field is an odd function under $z\rightarrow - z$. Apart from this, the electric-field evolution mirrors that of the magnetic field: for large $\sigma$, it is located close to the origin, with very small magnitude, while for smaller $\sigma$ it is peaked near the positions of the moving charges, and has a larger magnitude.

The left panel of Fig.\ \ref{fig:JBysigmaE} shows the time evolution of the charge diffusion current $qV^x$ (solid line) and $\sigma \mathcal{E}^x$ (dashed line) at $z=-0.006$ fm \footnote{We use a slight offset from the origin, because by symmetry the electric field $\mathcal{E}^x$ vanishes at that point.} for different values of $\sigma$, but the same $\tau_V=10^{-2}$ fm. We see that, although in both cases the charge diffusion current approaches its Navier-Stokes limit ($qV^x=\sigma \mathcal{E}^x$) at late times, the transient behavior at intermediate times depends on the damping ratio $\zeta_d$. 
For the blue curves, $\zeta_d \simeq 15.81$, which corresponds to the overdamped case. In this case, the charge diffusion current more or less follows the electric field without oscillations.
On the other hand, for the red curves, $\zeta_d = 1/2$, corresponding to the underdamped case. Here,
we observe a large phase shift between $qV^x$ and
$\sigma \mathcal{E}^x$, accompanied by an oscillatory behavior. 

The right panel of Fig.\ \ref{fig:JBysigmaE} shows the time evolution of the same quantities as in 
the left panel for two different values of $\tau_V$ keeping $\sigma=10^{-1}$ $\mathrm{fm}^{-1}$ fixed. In both cases, the parameters are chosen such that the evolution of the charge diffusion current is overdamped: for the red curves, $\zeta_d \simeq 1.58$, while for the blue curves $\zeta_d =5$. The two curves for $\sigma \mathcal{E}^x$ overlap, while the charge diffusion current for the larger relaxation time (red curve) takes longer to approach its Navier-Stokes value. It is also smaller in magnitude than for the smaller relaxation time, indicating an incomplete generation of the charge diffusion current. Such a phenomenon has also been recently seen in the transport approach of Ref.\ \cite{Wang:2021oqq}. This, in turn, will affect the rate of decay of the magnetic field in the medium as we will see next.
   \begin{figure}
    \centering
    \includegraphics[width=0.45\textwidth]{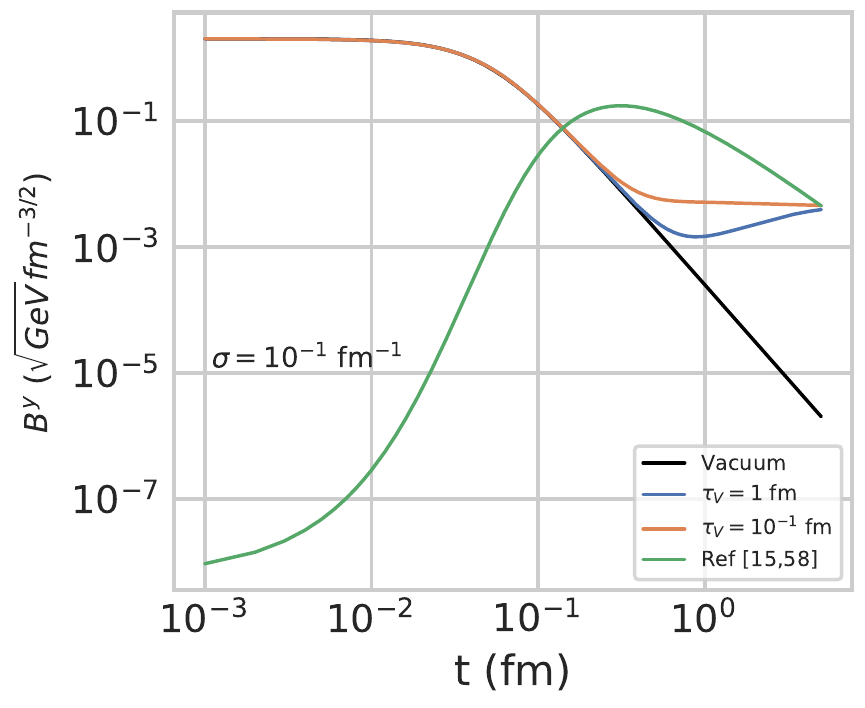}
    \includegraphics[width=0.47\textwidth]{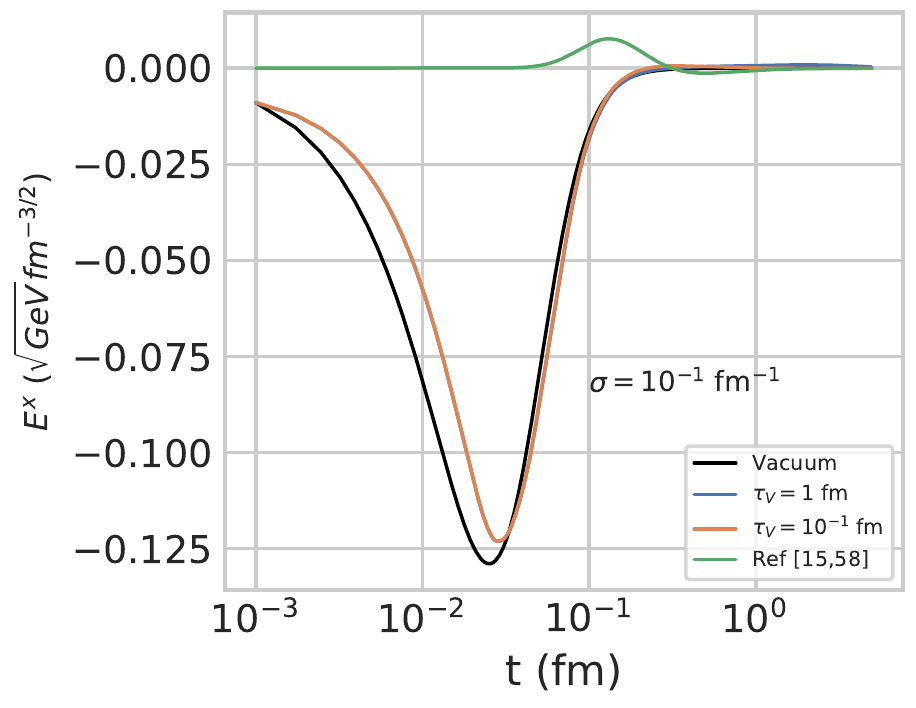}
    \caption{Left panel: The time evolution of $B^y$ at $z=-0.006$ fm. The solid black line corresponds to the solution in vacuum, i.e., for $\sigma=0\, \fm^{-1}$, whereas the solid green line corresponds to the solution obtained in Refs.\  \cite{Tuchin:2013ie,Gursoy:2014aka} for $\sigma=10^{-1} \fm^{-1}$. The solid orange and blue lines correspond to the numerical solution computed for the same value of $\sigma$, but using different values for the relaxation time $\tau_V$.  Right panel: Same as left panel but for $E^x$.}
    \label{fig:EMvst}
\end{figure}

   \begin{figure}
    \centering
    \includegraphics[width=0.45\textwidth]{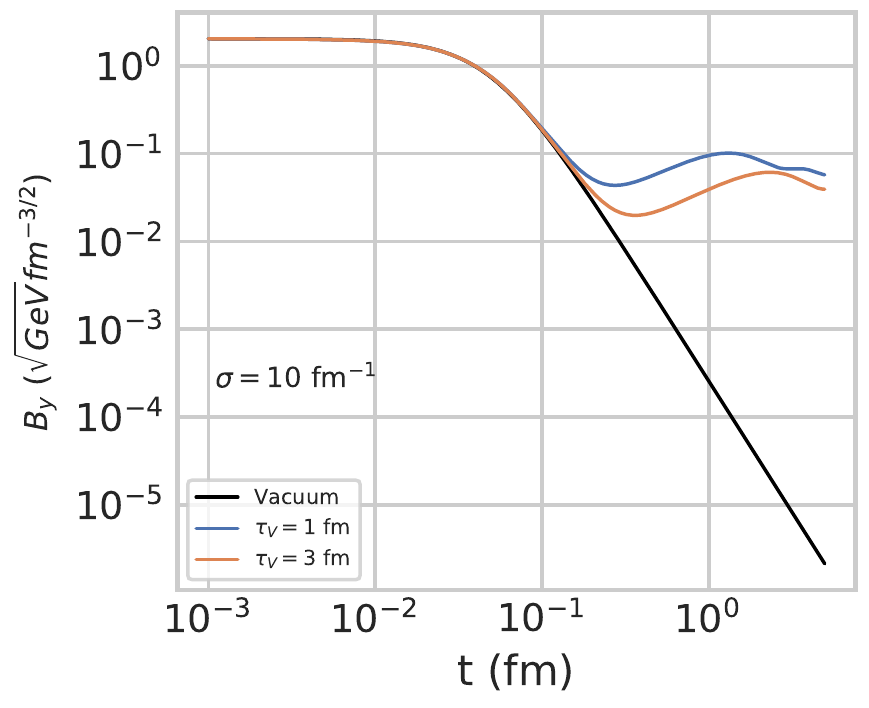}
    \includegraphics[width=0.46\textwidth]{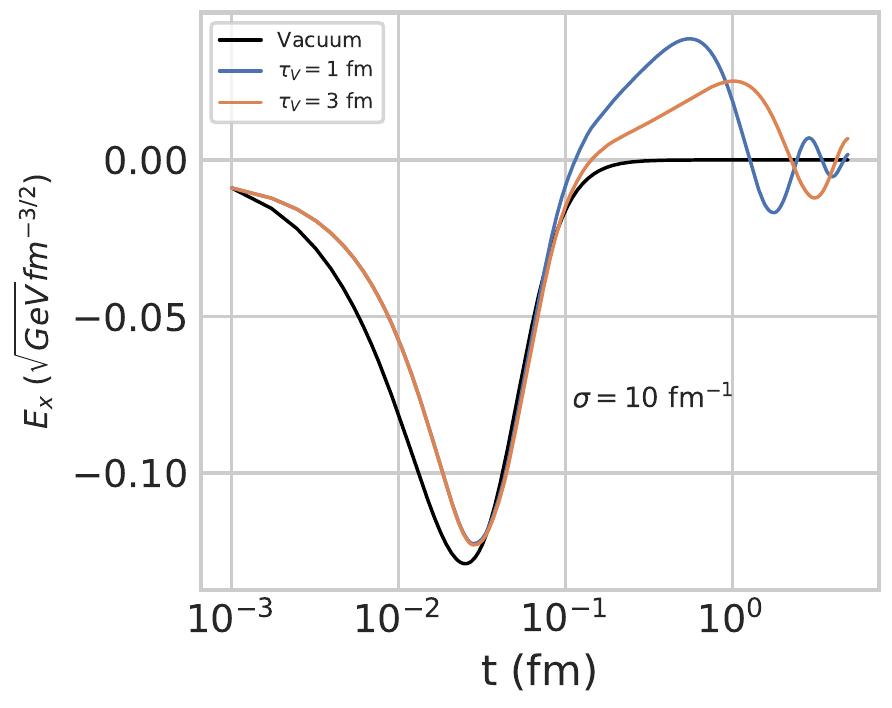}
    \caption{Left panel: The time evolution of $B^y$ at $z=-0.006$ fm. The solid black line corresponds to the solution in vacuum, i.e., for $\sigma=0\, \fm^{-1}$. The solid orange and blue lines correspond to the numerical solution computed for  $\sigma=10 \fm^{-1}$ , but for different values for the relaxation time $\tau_V$.  Right panel: Same as left panel but for $E^x$.}
    \label{fig:EMvst_UD}
\end{figure}

The left panel of Fig.\ \ref{fig:EMvst} shows the time evolution of $B^y$ evaluated at $z=-0.006\fm$ for $\sigma = 0 \, \fm^{-1}$ (solid black line) and $\sigma = 10^{-1} \fm^{-1}$, for various values of $\tau_V$. Due to the induced current generated in a medium with finite conductivity, the rate of decay of the magnetic field decreases (solid orange and blue lines). Moreover, as discussed in the previous paragraph, a longer relaxation time (the solid blue line as compared to the solid orange line) means an incomplete response of the charge diffusion current and hence leads to faster decay of the magnetic field at early times. However, at late times, for a larger relaxation time the induced current is still present, even though the electric field has decayed, see right panel of Fig.\ \ref{fig:JBysigmaE}. This, in turn, leads to a decrease in the decay rate of the magnetic field at late times. It can even reverse the decay, leading to an increase of the magnetic field, as observed for the solid blue line in the left panel of Fig.\ \ref{fig:EMvst}. The solid green line shows the solution obtained by solving Maxwell's equation in a conducting medium shown previously in Refs.\ \cite{Tuchin:2013ie,Gursoy:2014aka}, using the Navier-Stokes form (\ref{eq:Ohmslawacausal}) of Ohm's law. The right panel of Fig.\ \ref{fig:EMvst} shows the time evolution of $E^x$ evaluated at $z=-0.006\fm$, for the same parameters as in the left panel. Here, the solid blue and orange lines overlap. 

Notice that, as soon as one introduces a nonzero relaxation time $\tau_V$, no matter how small, the initial evolution of the electromagnetic fields must follow that of the vacuum, since the back-reaction of the medium needs some finite amount of time to build up.
The solution of Refs.\ \cite{Tuchin:2013ie,Gursoy:2014aka} does not account for this, since it uses the instantaneous Navier-Stokes form (\ref{eq:Ohmslawacausal}) of Ohm's law. The strong suppression of this solution as compared to the vacuum one is due to specific form of the solution, cf., e.g.,  Eq.\ (2.3) in Ref.\ \cite{Gursoy:2014aka}, which features an exponential behavior $\sim \exp (- \sigma b\, \sinh Y_{\text{bm}}/4) \sim
\exp( - 130 \sigma \, \fm)$ for
RHIC energies ($Y_{\text{bm}} = 5$) and an impact parameter $b=7\, \fm$. Thus, for  $\sigma = 10^{-1} \fm^{-1}$, the initial electromagnetic fields will be smaller by several orders of magnitude as compared to the vacuum solution, as one observes in the left panel of Fig.\ \ref{fig:EMvst}.

In Fig.\ \ref{fig:EMvst_UD} we show the solution for the underdamped case, i.e., for $\sigma = 10\, \fm^{-1}$ and $\tau_V = 1$ and 3 fm, respectively. While the evolution is similar as in Fig.\ \ref{fig:EMvst}, at late times one observes oscillations typical for the underdamped case.

   \begin{figure}
    \centering
    \includegraphics[width=0.45\textwidth]{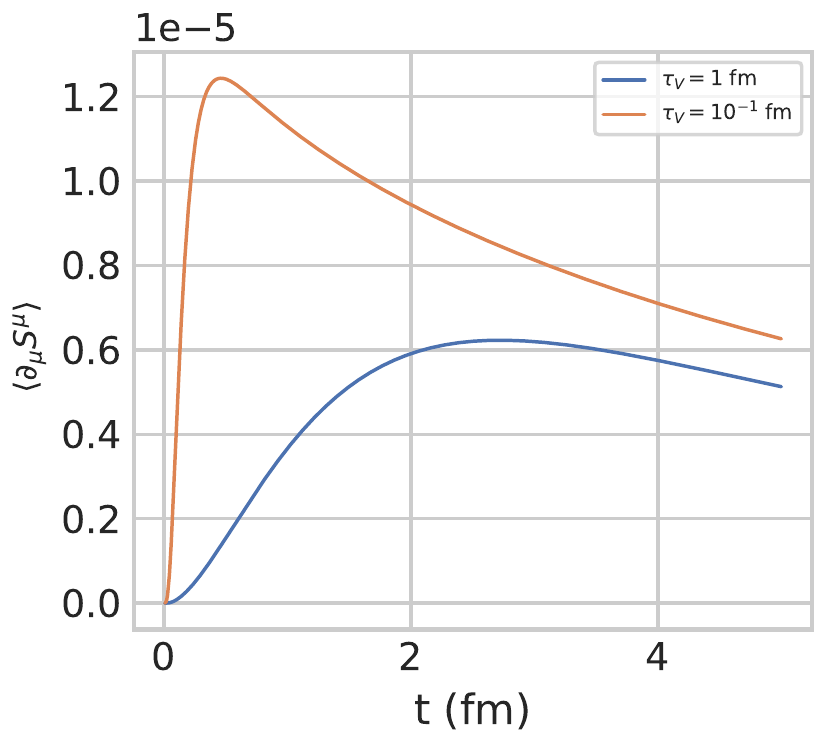}
    \includegraphics[width=0.463\textwidth]{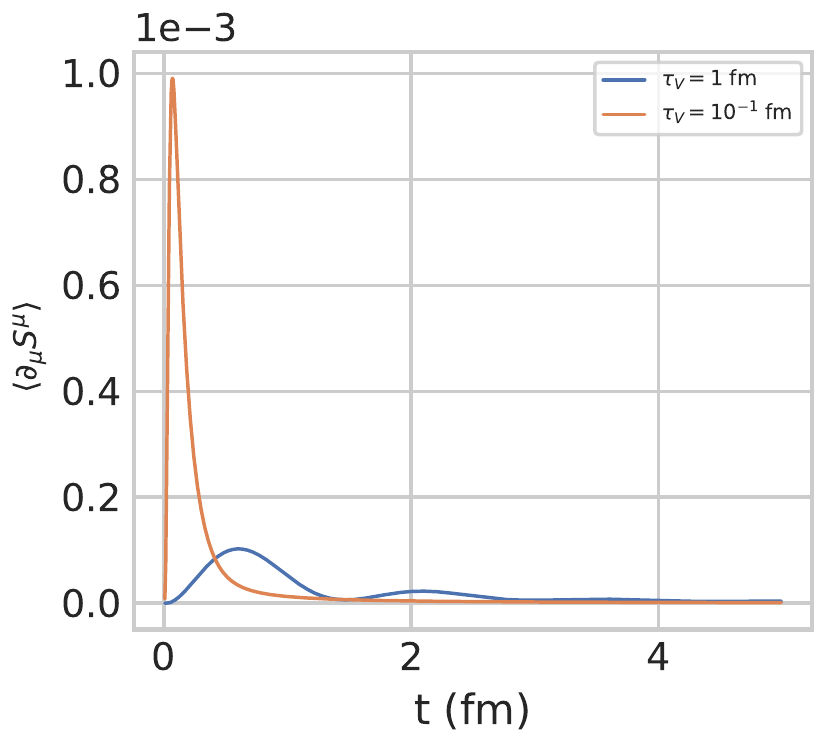}
    \caption{Left panel: The time evolution of grid-averaged entropy production rate scaled by the pressure, $\langle \partial_\mu S^\mu \rangle$, as a function of time for a finite constant conductivity $\sigma=10^{-1} \fm ^{-1}$. Solid orange and blue lines correspond to the numerical solution using different relaxation times $\tau_V$.  Right panel: Same as left panel but for  $\sigma=10\, \fm^{-1}$.}
    \label{fig:Entropyvt}
\end{figure}

Next, we also compute the entropy production due to Ohmic dissipation. The rate of entropy production is also proportional to the rate of decrease of electromagnetic energy in the system \cite{schwinger2019classical}. The entropy current $S^\mu$  is given as 
\begin{equation}
S^\mu=P\beta^\mu+T^{\mu\nu}_f\beta_\nu-\frac{\alpha}{q} J^\mu_f +\frac{\delta}{2}u^\mu V^\nu V_{\nu}+\mathcal{O}_3\;,
\end{equation}
where $\mathcal{O}_3$ denotes terms of third order or higher in the dissipative currents. The coefficient $\delta$ is a function of the temperature and chemical potential of the equilibrium state and can only be obtained by matching this expansion with
the underlying microscopic theory. Here we keep it as a free parameter.

Taking the divergence of the above equation and using the conservation laws (\ref{eq:partialTmunu}) we arrive at the following expression 
\begin{equation}
    \partial_\mu S^\mu=V_\mu\left[-\frac{q}{T}\mathcal{E}^\mu-\nabla^\mu\alpha +\delta\,\dot{V}^{\mu}+V^{\mu}\partial_\nu\left(\frac{\delta}{2}u^\nu\right) \right]\;.
\end{equation}
The fourth term on the right-hand side can be neglected if we restrict the
discussion to the simplest
possible case, as has already been done in Eq.~(\ref{eq:Ohmslaw}). Enforcing the positivity of the second law by requiring
\begin{equation} \label{eq:2ndlaw}
    \partial_\mu S^\mu=-\frac{q^2}{\sigma T}V^\mu V_{\mu}
\end{equation}
 leads to the evolution equation (\ref{eq:Ohmslaw}) for the charge-diffusion current. The coefficient $\delta$ can then be identified with $\delta=q^2\tau_V/(\sigma T)$. The left panel of Fig.\ \ref{fig:Entropyvt} shows the grid-averaged entropy production rate,
\begin{equation}
     \langle \partial_\mu S^\mu\rangle\coloneqq\frac{\int d^3x\, \partial_\mu S^\mu}{\int P d^3 x} =\frac{\int dz\, \partial_\mu S^\mu}{\int P dz} 
\end{equation}
 as a function of time with constant conductivity $\sigma=10^{-1}\, \fm^{-1}$. The spatial ($z$-) integration is taken over the whole grid.
Notice that in case of a longer relaxation time (solid blue line), the rate of entropy production is delayed as compared to the case of a shorter relaxation time (solid orange line). The right panel of Fig.\ \ref{fig:Entropyvt}  shows the average entropy production rate for a larger value of the conductivity, $\sigma=10 \, \fm^{-1}$. For such a large conductivity, a value of $\tau_V=1\, \fm$ for the relaxation time renders the charge diffusion current underdamped (as then
$\zeta_d \simeq 0.158$). Consequently, the entropy production rate oscillates (solid blue line). If we reduce the relaxation time to $\tau_V=10^{-2}\fm$, for $\sigma=10 \, \fm^{-1}$ the charge diffusion current is again in the overdamped region and the entropy production rate no longer oscillates (solid orange line). Qualitatively, this case looks like the cases shown in the left panel of Fig.\ \ref{fig:Entropyvt}, however, since
$\tau_V$ is at least an order of magnitude smaller, the decrease of the entropy production rate occurs on a much smaller timescale as well. Naturally, as expected from Eq.\ (\ref{eq:2ndlaw}), the entropy production is never negative.

Finally, we remark that, because of a finite Poynting vector $\sim \boldsymbol{E} \times \boldsymbol{B}$, the electromagnetic field also accelerates the fluid, giving rise to a nonzero fluid velocity $v^z$.
However, we do not discuss this here but rather move to the more physical case of a finite initial expansion rate of the fluid, where we investigate these effects in more detail.

\subsection{Initially expanding fluid}

We now consider a system which is initially expanding according to Bjorken's scaling-flow scenario \cite{Bjorken:1982qr}, i.e.,
with initial flow velocity in $z$-direction $v^z\equiv z/t\equiv \tanh \eta$, where 
$\eta \coloneqq \mathrm{Artanh}\, v^z$ is the space-time rapidity. The transverse velocity is $\boldsymbol{v}_\bot \equiv 0$.
The proper time of a fluid element moving with
$v^z = \tanh \eta$ is denoted by $\tau$. For such a flow profile, it is usually convenient to replace the standard Minkowski coordinates ${x}^\mu = (t,x,y,z)$ by Milne coordinates $\tilde{x}^\mu = (\tau,x,y,\eta)$, with metric tensor $g_{\mu\nu}=\mathrm{diag}(1,-1,-1,-\tau^2)$. In these coordinates, the equations of motion \eqref{Eq:systemofeqn} acquire some additional source terms, for more details see App.~\ref{App:Milnecord}. 

We consider the same colliding system
as in Sec.\ \ref{sec:IIIA}, i.e., a Au-Au collision at $\sqrt{s_{NN}}=200$ GeV at impact parameter $b=10$ fm. We initialize the
evolution at $\tau_0 =0.1\, \fm$ with   uniform energy density $\varepsilon=13.33$ $\mathrm{GeV/fm}^3$. Since we also want to investigate the case of a temperature-dependent conductivity, we need to relate the temperature to a fluid-dynamical quantity, e.g. the energy density. Assuming the particles in our system to be massless spin-1/2 fermions, this relationship is given by $\varepsilon \equiv  7 \pi^2 T^4/60$. The Faraday tensor $\tilde{F}^{\mu\nu}$ in Milne coordinates at initial time $\tau_0 = 0.1 \, \fm$ can be computed via a coordinate transformation,
\begin{align}\label{eq:CoordTransform}
      \tilde{F}^{\mu\nu}(\tau_0, \boldsymbol{x}_\bot, \eta)=\frac{\partial \tilde x^\mu}{\partial x^\rho}\frac{\partial \tilde x^\nu}{\partial x^\sigma}F^{\rho\sigma}(\tau_0 \cosh \eta,\boldsymbol{x}_\bot, \tau_0 \sinh \eta)\;,
\end{align}
where the initial Faraday tensor $F^{\rho\sigma}$ in Minkowski coordinates is calculated using Eqs.\ (\ref{eq:Intialemfields1}) and (\ref{eq:Intialemfields2}). Since we assume homogeneity in the transverse direction, it is sufficient to consider all fields at $\boldsymbol{x}_\bot=0$. The further
evolution of the electromagnetic fields is then determined by solving Maxwell's equations in Milne coordinates.

   \begin{figure}
    \centering
    \includegraphics[width=0.45\textwidth]{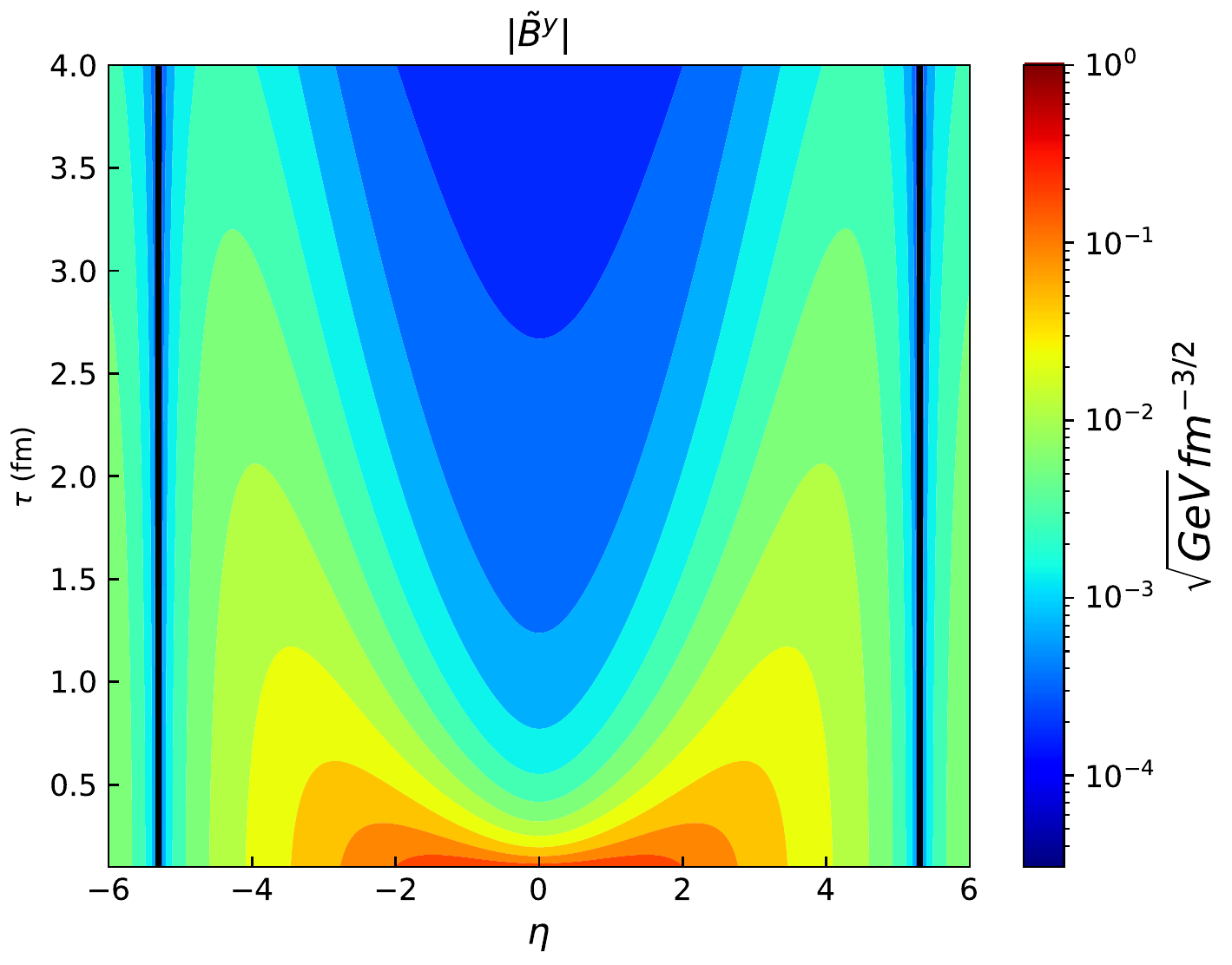}
    \includegraphics[width=0.45\textwidth]{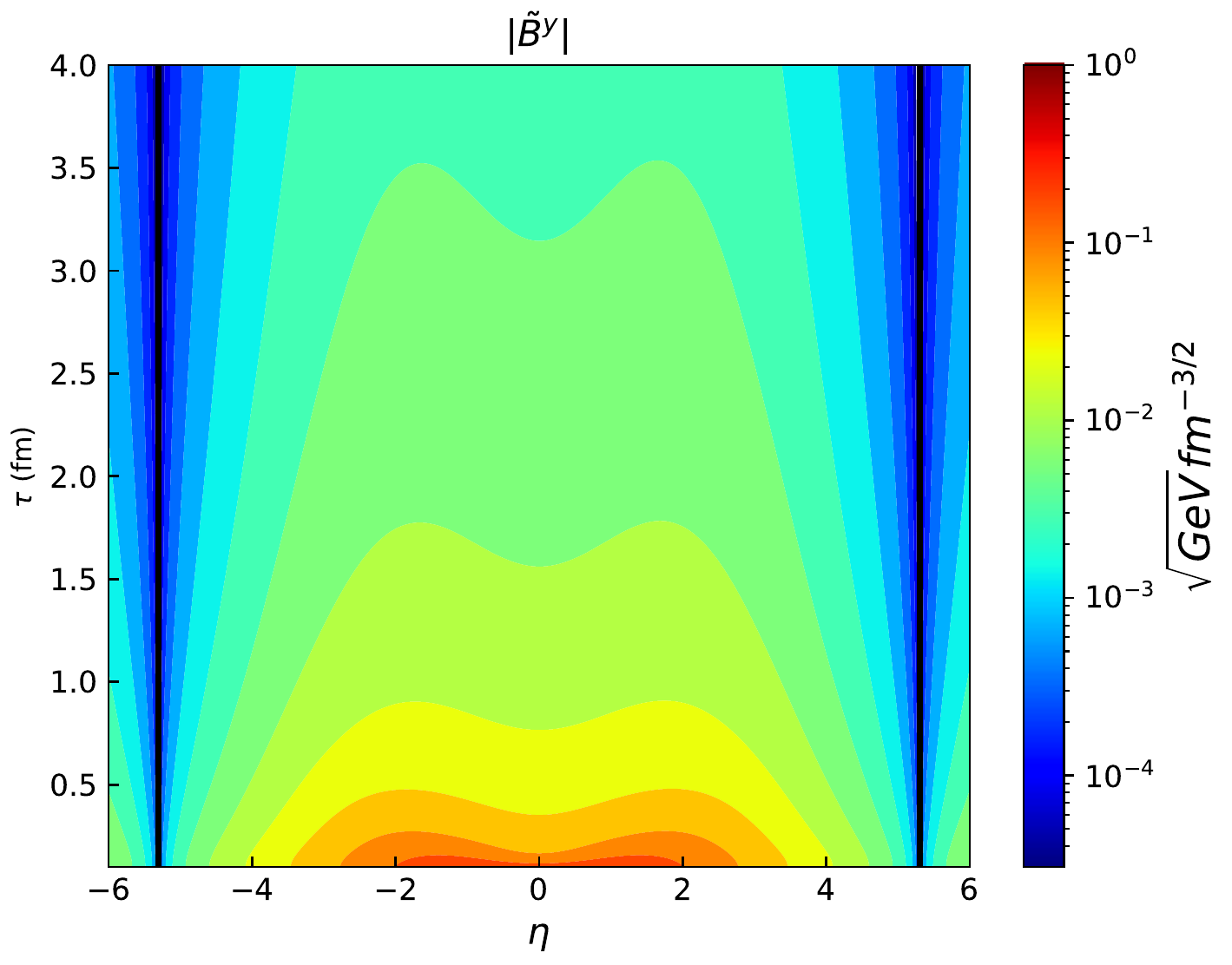}
     \includegraphics[width=0.45\textwidth]{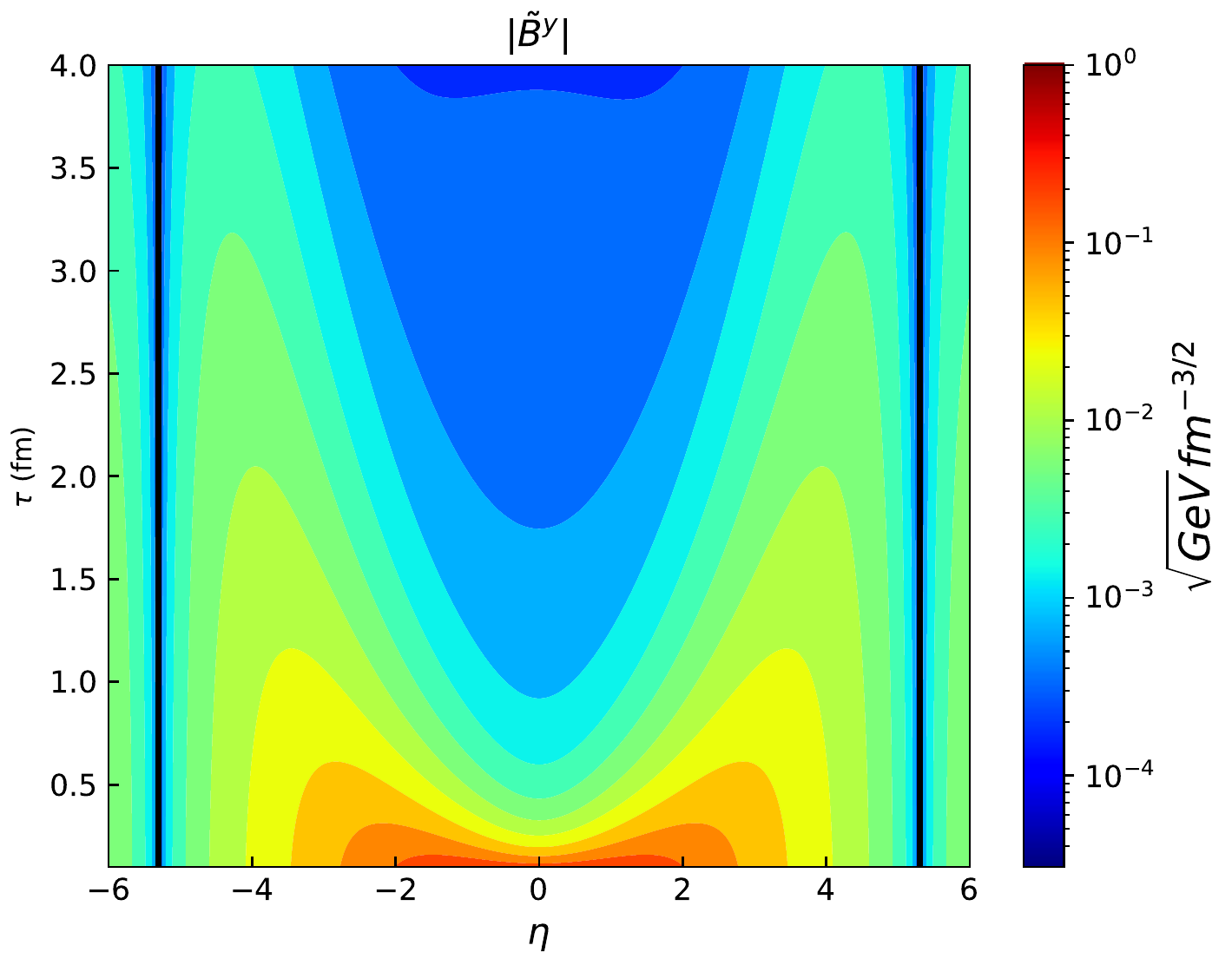}
    \caption{Top: Contour plot of 
    $|\tilde{B}^y|$ (in units of $\sqrt{\GeV}\fm^{-3/2}$) in the $(\tau, \eta)-$plane for $\sigma=10^{-1}\fm^{-1}$ (left panel) and $\sigma=10\,\fm^{-1}$ (right panel). Bottom: Same as above, but for a temperature-dependent conductivity $\sigma/T=8 \pi \alpha_{\mathrm{EM}}/3$. The relaxation time is chosen as $\tau_V=10^{-2}$ fm.}
    \label{fig:MilneByvst}
\end{figure}

Figure \ref{fig:MilneByvst} shows a
contour plot of $|\tilde{B}^y|$ in the $(\tau,\eta)$ plane for various values of the conductivity. The black vertical lines
indicate beam rapidity $\pm Y_{\mathrm{bm}}$. One feature to note is that the magnetic field is positive inside the region $|\eta|<Y_{\mathrm{bm}}$ and negative outside. For a small value of $\sigma$ the peak in the magnetic field remains close to the rapidities
$\pm Y_{\mathrm{bm}}$ of the initial charges, with diffusive tails extending towards the mid-rapidity region, cf.\ also Fig.\ \ref{fig:NonexpandingEmfields}. On the other hand, for a larger value of $\sigma$, the magnetic field remains at mid-rapidity as a consequence of the frozen-flux theorem,
cf.\ Fig.\ \ref{fig:NonexpandingEmfields}. For a temperature-dependent conductivity, which in our case we take to be $\sigma/T= 8 \pi \alpha_{\mathrm{EM}}/3 \simeq 0.06$ \cite{Ding:2010ga,Amato:2013naa}, the dynamics is non-trivial since the system transits from a conducting to an insulating medium as it cools down with time. Nevertheless, for such a small value of $\sigma/T$, the evolution is rather similar to the case with a constant small $\sigma$ shown in the upper left panel of Fig.\ \ref{fig:MilneByvst}. We also note that, as the conductivity $\sigma$ increases, the magnetic field at late times approaches the ideal-MHD scaling limit for Bjorken flow,
i.e., at fixed $\eta$, $\tilde{B}^y \sim 1/\tau$, as found previously in Ref.\ \cite{Roy:2015kma}. 

\begin{figure}
    \centering
    \includegraphics[width=0.45\textwidth]{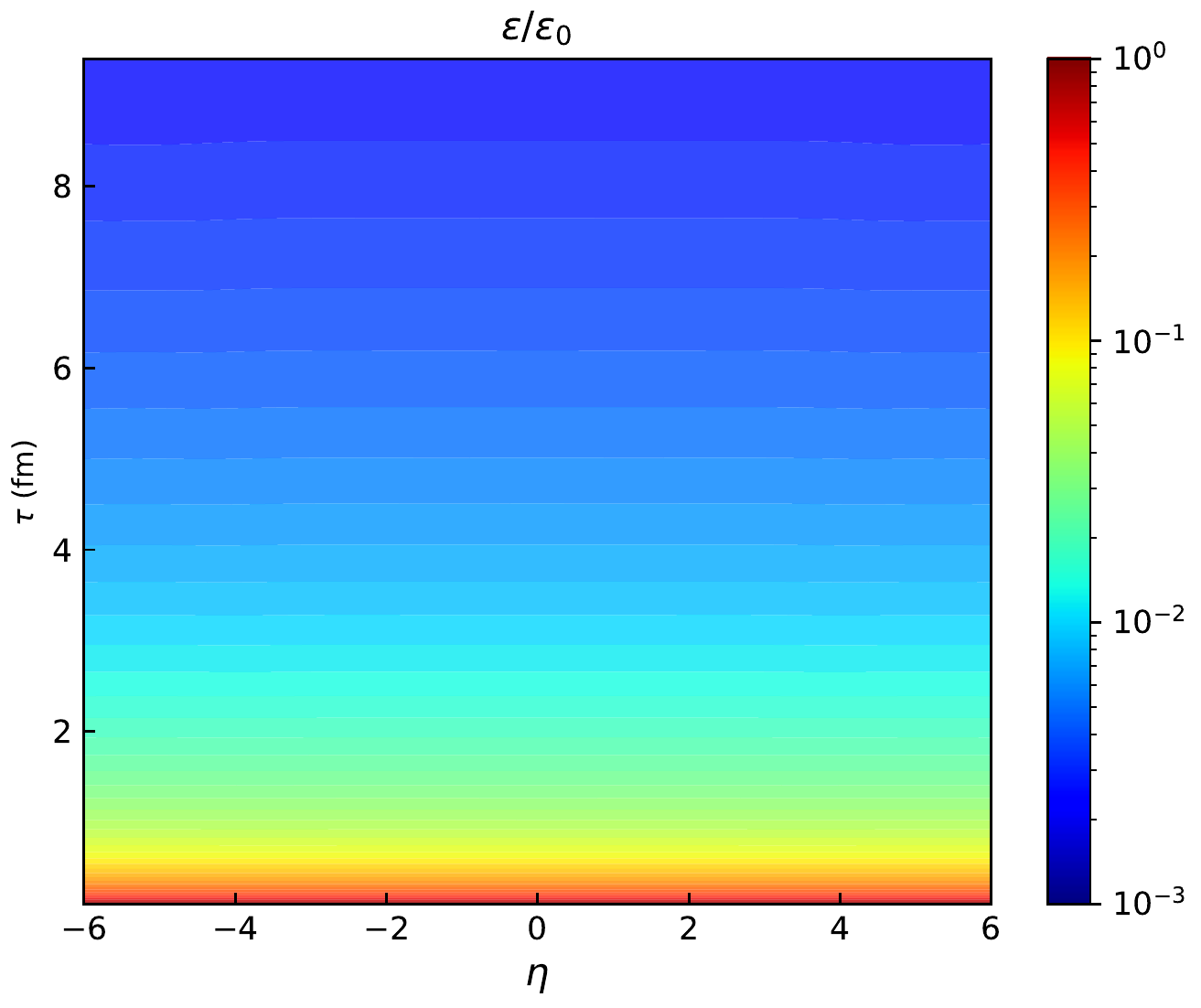}
    \includegraphics[width=0.45\textwidth]{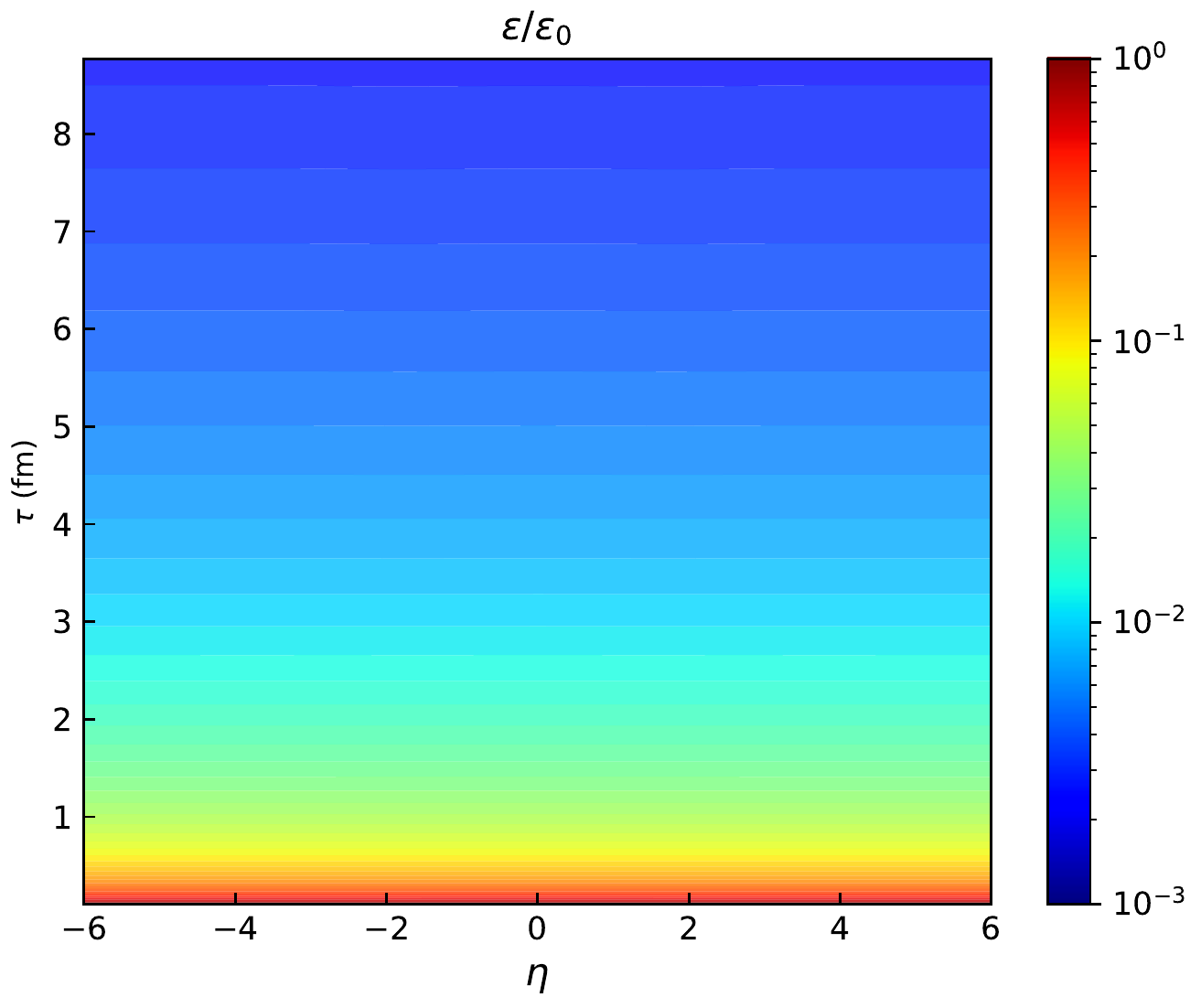}
     \includegraphics[width=0.45\textwidth]{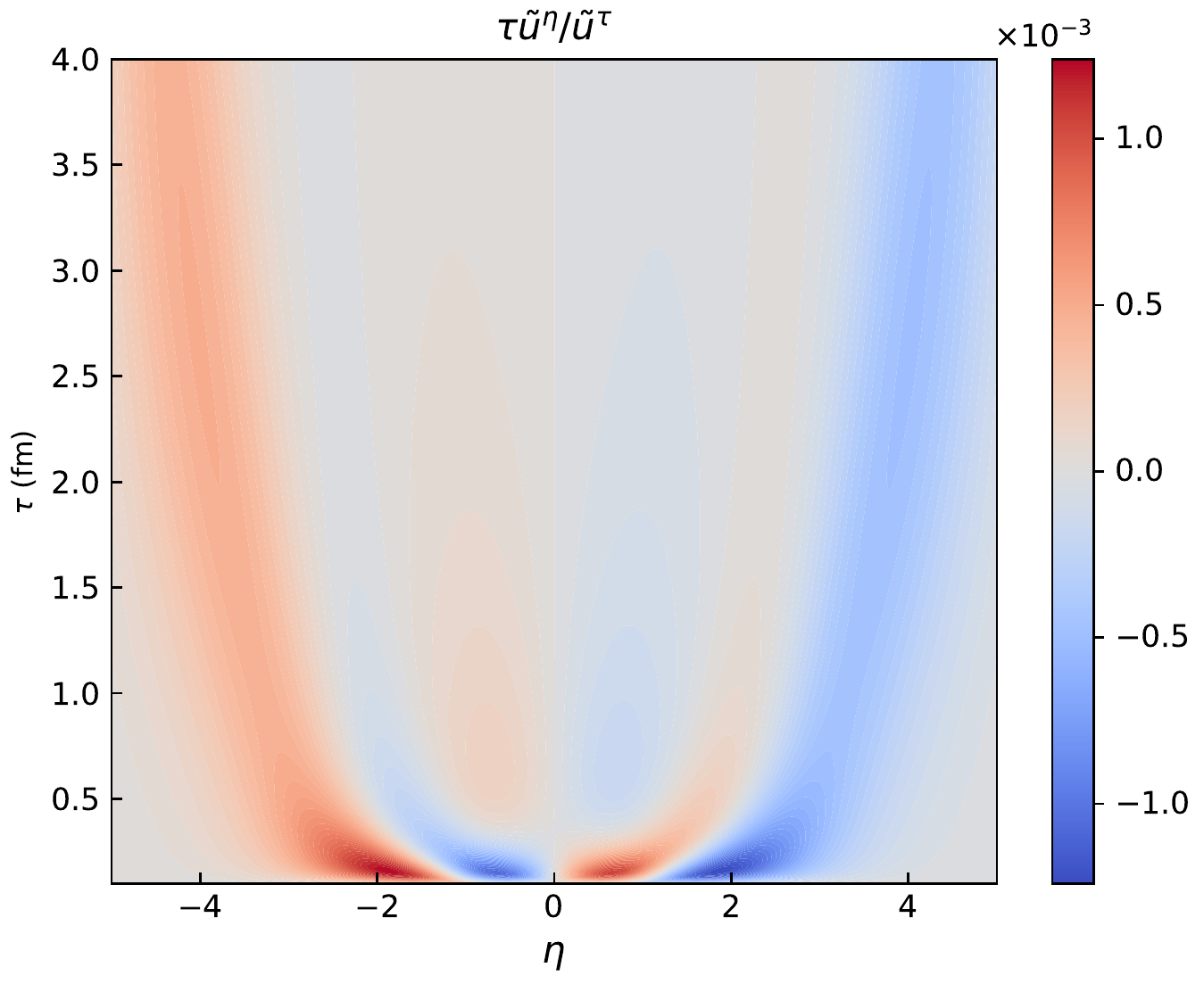}
     \includegraphics[width=0.45\textwidth]{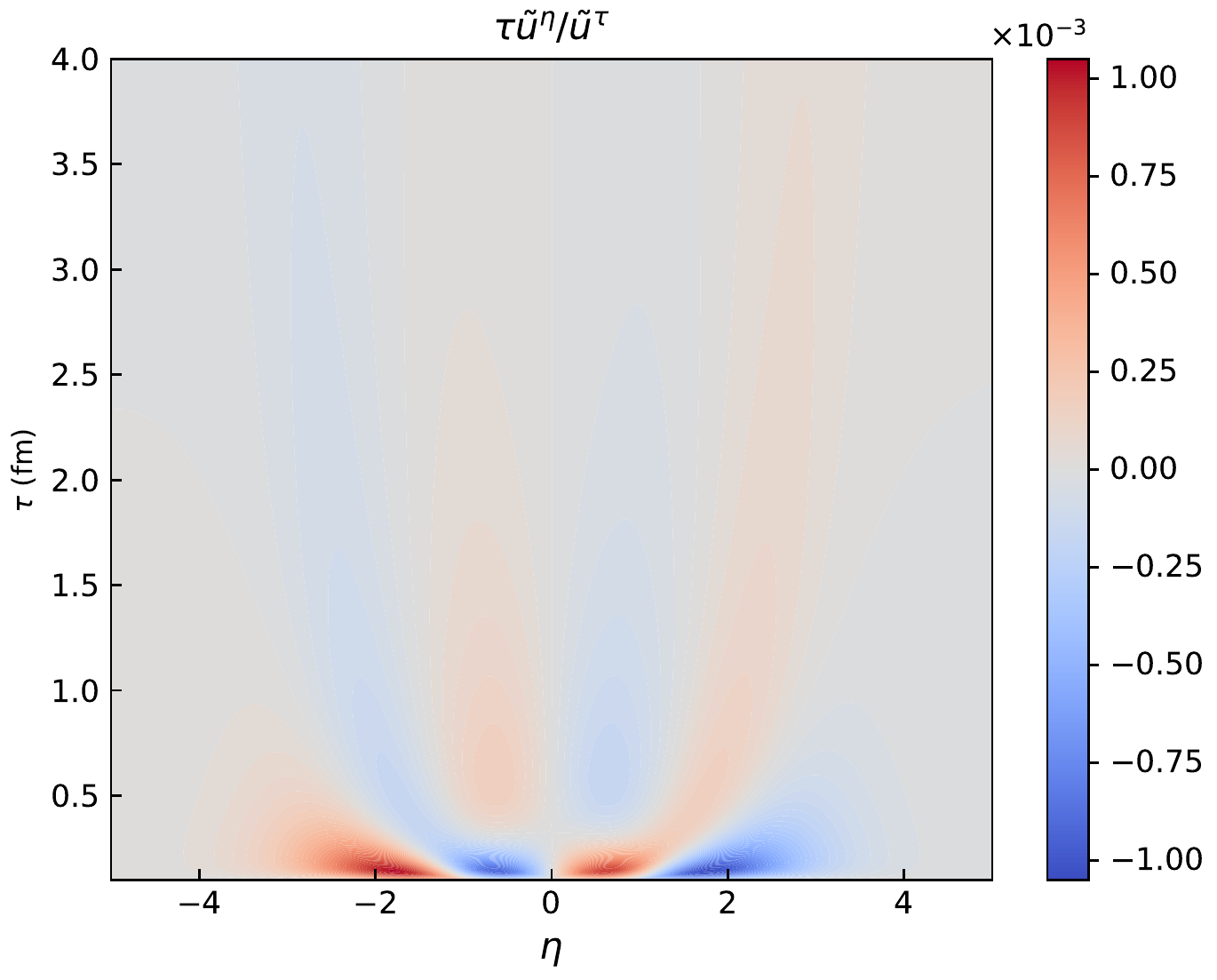}
    \caption{Top row: Contour plot of the energy density $\varepsilon$ in the $(\tau,\eta)-$plane for $\sigma=10^{-1} \fm^{-1}$ (left panel) and $\sigma=10 \, \fm^{-1}$ (right panel). Bottom row: Same as top row, but for the fluid velocity in $\eta$-direction, $\tau \tilde{u}^\eta/\tilde{u}^\tau$. The relaxation time is chosen as 
    $\tau_V=10^{-2}$ fm.}
    \label{fig:Milneendtyandvz}
\end{figure}

To study the back-reaction of the electromagnetic fields on the fluid, we consider the energy density and the fluid velocity. For pure Bjorken flow, $\varepsilon$ should remain constant as a function of $\eta$ and decay with $\tau$
as $\varepsilon \sim \tau^{-4/3}$. Moreover, in pure Bjorken flow $\tilde{u}^\mu = (1,0,0,0)$. The back-reaction should then in principle be visible in violations of this behavior. Nevertheless, in the ideal-MHD limit and in the absence of an external charge current, energy and momentum of the fluid are separately conserved (see, e.g., Ref.\  \cite{Denicol:2018rbw}), i.e.,
\begin{equation}
    \partial_\mu T_f^{\mu\nu}=0 \;.
\end{equation}
This is because the electric field goes to zero in the ideal-MHD limit and the magnetic field influences the dynamics of the fluid only by coupling to the dissipative part of the charge current $qV^\mu$. However, without dissipation, the flow is adiabatic and the dynamics of the fluid remains unaffected by the magnetic field. Therefore, for large values of $\sigma$ we also do not expect any significant deviations from the Bjorken-flow scenario, and a notable back-reaction on the fluid should only be observable for small values of $\sigma$.

These expectations are borne out by our numerical calculations. In Fig.\ \ref{fig:Milneendtyandvz} (top row) we show contour plots of the energy density scaled
by its value at $\tau_0$, $\varepsilon/\varepsilon_0$, in the $(\tau,\eta)-$plane for a small value of $\sigma=10^{-1} \fm^{-1}$ (left panel) and a large value of $\sigma=10\, \fm^{-1}$ (right panel), respectively. For both large and small values of the conductivity, there is no visible breaking of boost invariance in the energy density (upper row).
We have also checked the scaling of the energy density with $\tau$ around mid-rapidity and find that it closely follows the ideal Bjorken-scaling law $\sim \tau^{-4/3}$.

Considering the
fluid four-velocity, we show its
$\eta$-component (multiplied by $\tau$ in order to make it dimensionless, and divided by its $\tau$-component) in the bottom row of Fig.\ \ref{fig:Milneendtyandvz}.
Deviations from Bjorken flow are visible by the generation of nonzero values of $\tau \tilde{u}^\eta/\tilde{u}^\tau$. These deviations are larger (smaller) for smaller (larger) values of $\sigma$, which is evident from comparing the magnitude of 
$\tau \tilde{u}^\eta/\tilde{u}^\tau$ in the left and the right panels of the bottom row of
Fig.\ \ref{fig:Milneendtyandvz}. 

\begin{figure}
    \centering
    \includegraphics[width=0.48\textwidth]{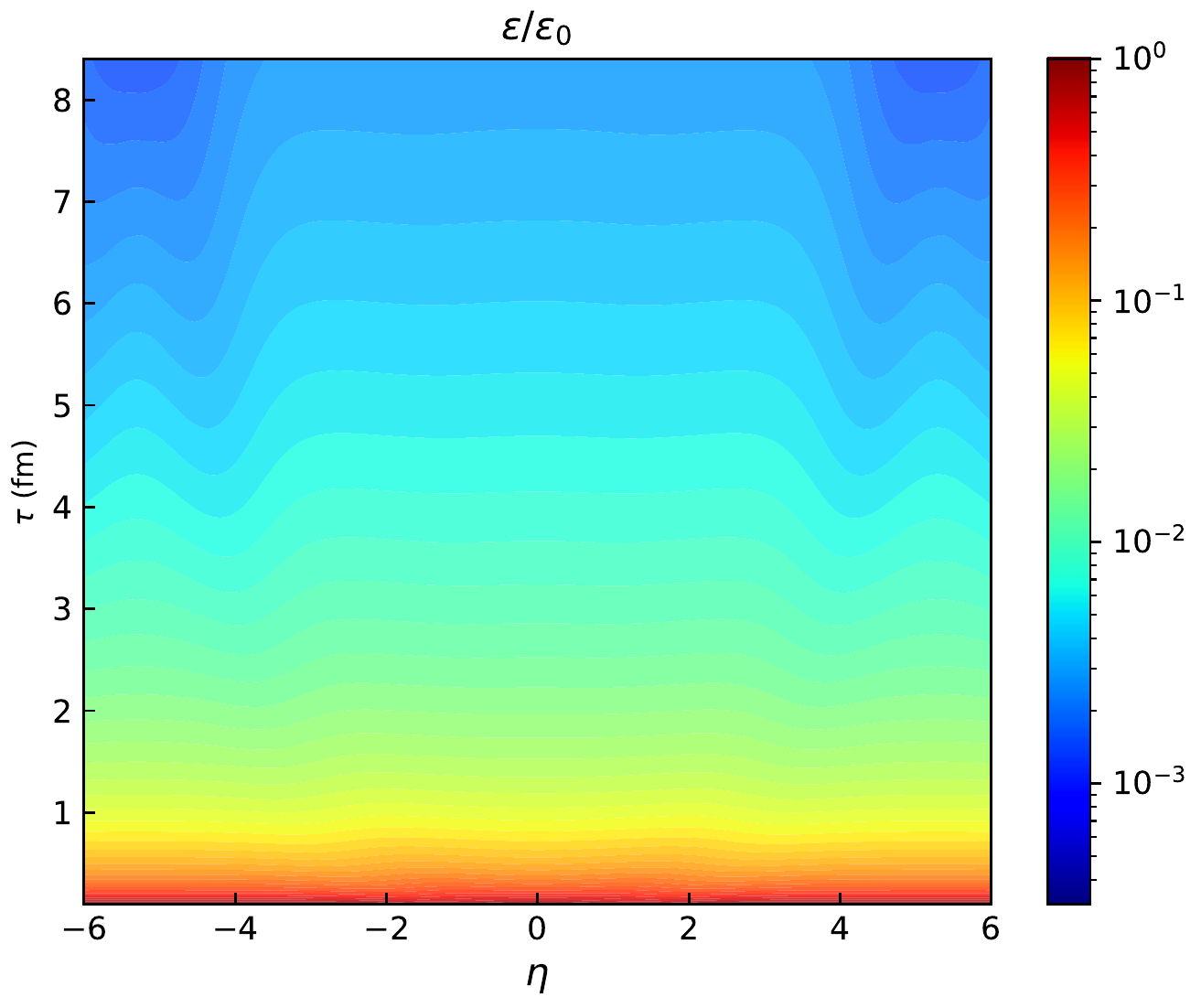}
     \includegraphics[width=0.48\textwidth]{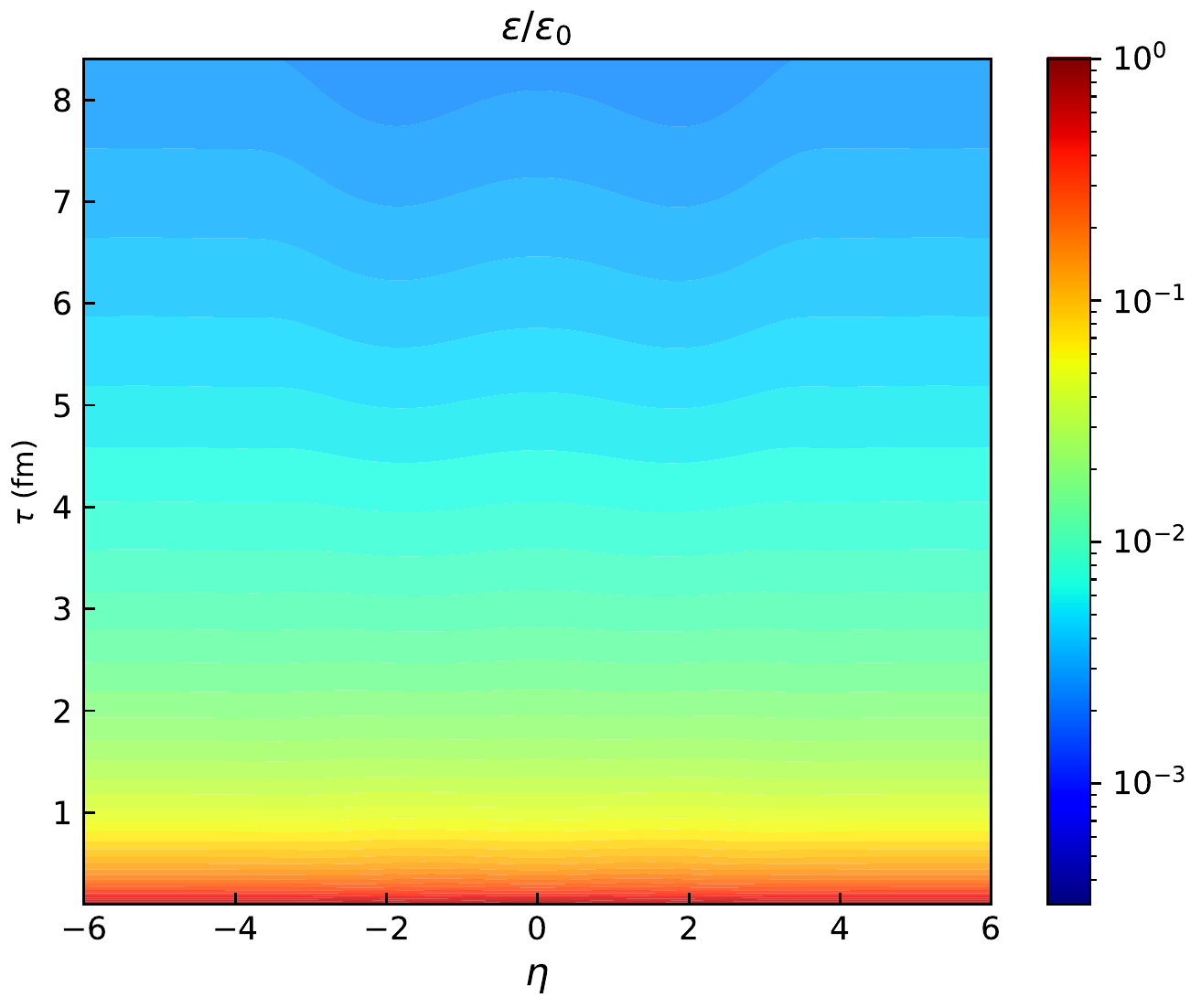}
    \includegraphics[width=0.50\textwidth]
    {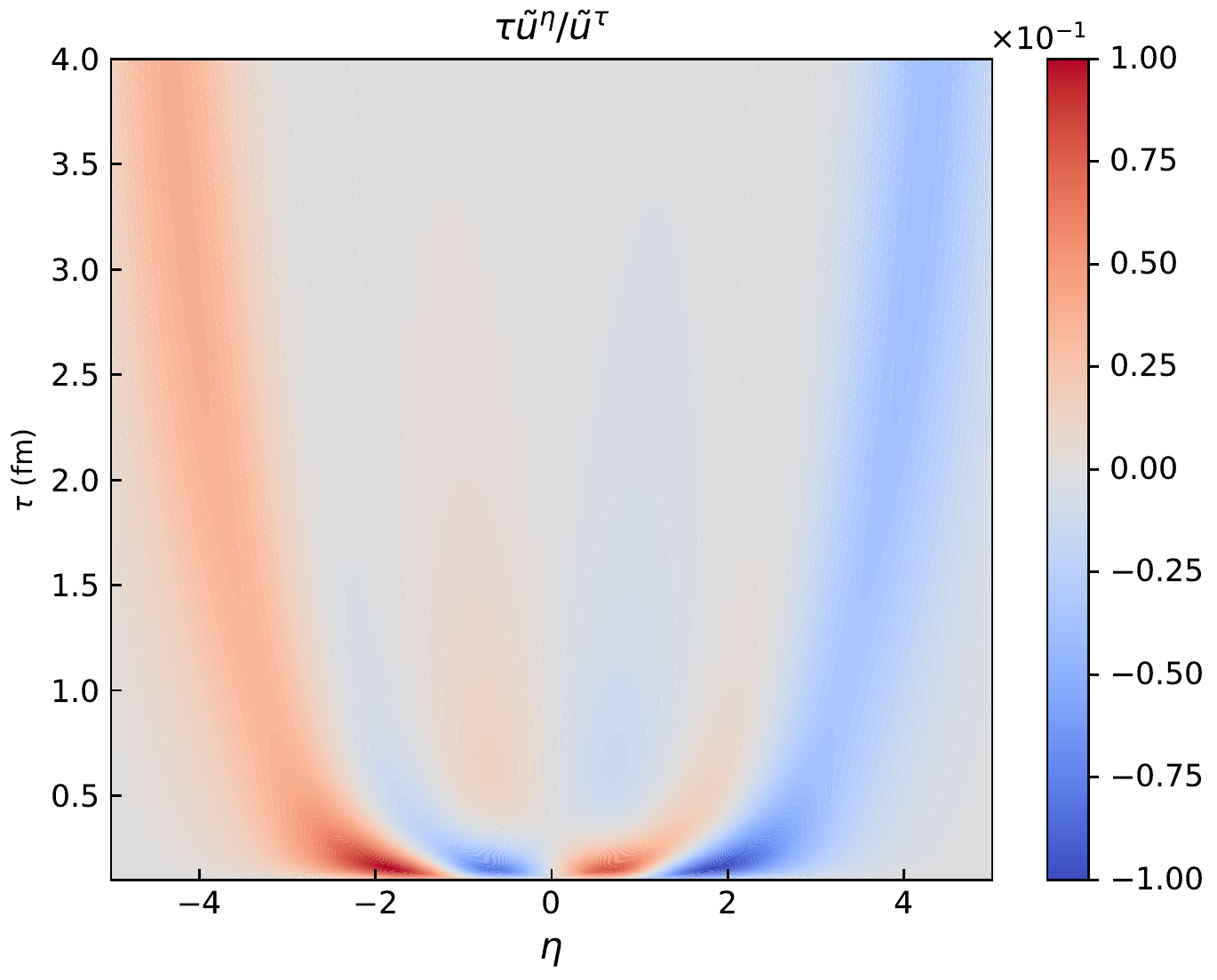}
     \includegraphics[width=0.48\textwidth]{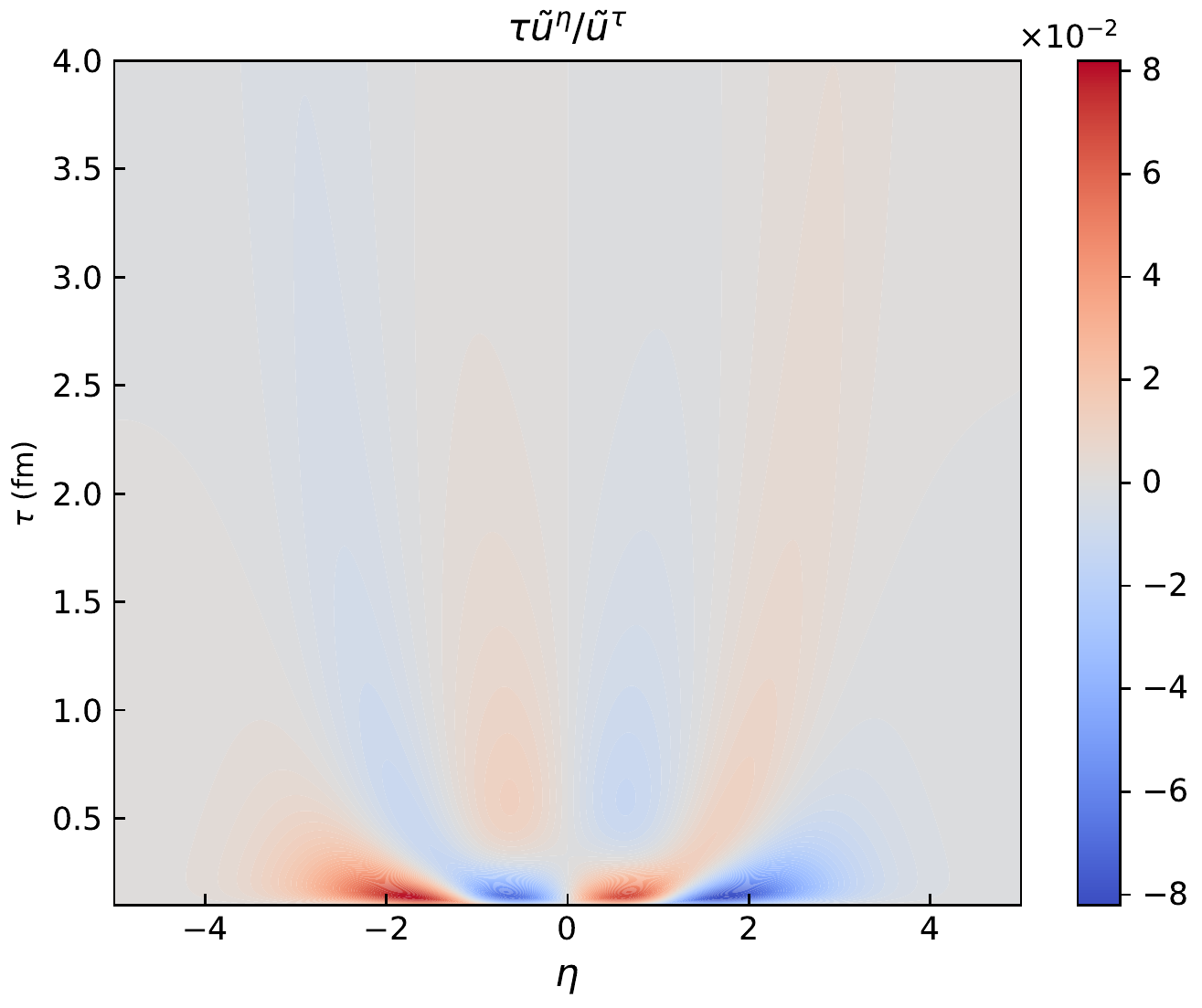}
    \caption{Top row: Contour plot of the energy density $\varepsilon$ in the $(\tau,\eta)-$plane for $\sigma=10^{-1} \fm^{-1}$ (left panel) and $\sigma=10 \, \fm^{-1}$ (right panel). Bottom row: Same as top row, but for the fluid velocity $\tau \tilde{u}^\eta/\tilde{u}^\tau$. The relaxation time is chosen as
    $\tau_V=10^{-2}$ fm and inverse plasma $\beta^{-1}=8$.}
    \label{fig:MilneendtyandvzInvbeta}
\end{figure}

We observe from Fig.\ \ref{fig:Milneendtyandvz} that the back-reaction of the magnetic field onto the fluid is quantitatively very small. The reason for this is that the initial value of the field energy density is small compared to that of the fluid. This is quantified by the so-called inverse plasma $\beta$-parameter.
In this context, we choose to define it as $\beta^{-1} \coloneqq B_0^2/(2P_0)$, where
$B_0^2/2$ is the initial magnetic pressure and $P_0$ the initial pressure of the fluid at $\boldsymbol{x}_\bot = z =0$. For the
parameters in Fig.\ \ref{fig:Milneendtyandvz} we have
$\beta^{-1} = 0.15$. If we artificially scale up the initial value of the electromagnetic fields such that $\beta^{-1} = 8$, we obtain a different picture, cf.\ Fig.\ \ref{fig:MilneendtyandvzInvbeta}. Such large values are not entirely unrealistic, in particular in the outer layers of the collision zone. As one can see from Fig.\ \ref{fig:MilneendtyandvzInvbeta} for  $\beta^{-1}=8$ the breaking of boost invariance increases substantially as compared to Fig.\ \ref{fig:Milneendtyandvz}. Now there are also visible deviations from Bjorken flow in the energy density, both for large and small values of the conductivity, and the fluid velocity can be two orders of magnitude larger than in the previous case. As one observes, the effect is larger for a smaller value of the conductivity, cf.\ left and right columns of Fig.\ \ref{fig:MilneendtyandvzInvbeta}.

\section{Conclusion and outlook}\label{sec:ConcandOut}

We have numerically solved the equations of motion of relativistic resistive second-order dissipative magnetohydrodynamics in a dimensionally reduced set-up, which is nevertheless relevant for heavy-ion collisions. The present framework is self-consistent, meaning it includes the back-reaction from electromagnetic fields to the fluid and vice versa. The fluid couples to the electromagnetic field via the charge diffusion, which is usually treated in the Navier-Stokes form of Ohm's law. As this introduces acausalities and instabilities in the relativistic case, we generalized the constituent relation for the charge diffusion current to an equation of motion including a relaxation term of Maxwell-Cattaneo form.
From the discussion of ordinary second-order dissipative fluid dynamics,  such a constituent relation is known to in principle restore causality and stability. 
The equation of motion for the charge diffusion current can be derived from kinetic theory using the Boltzmann-Vlasov equation
\cite{Denicol:2018rbw, Denicol:2019iyh}. We showed that, in the local rest frame of the fluid, such an equation, coupled to Maxwell's equations, can be recast into the equation of motion for a damped harmonic oscillator. The set of magnetohydrodynamical equations of motion features source terms which are stiff when the conductivity is large (e.g.\ in the ideal-conducting limit). Therefore, the
solution requires a numerical method which is able to handle such stiff source terms. Here, we employed an Implicit-Explicit Runge-Kutta (IMEX) method and checked its validity for various non-trivial test cases. 
We then applied our reduced magnetohydrodynamical set-up to two scenarios: one where the fluid is initially at rest and one where it initially expands longitudinally according to the Bjorken scaling flow.
Depending on the choices for conductivity and relaxation time, we found qualitatively different solutions. As expected, if the relaxation time is large, the evolution of the charge diffusion current does not follow that of the electric field, but may considerably lag behind in time and may not reach the full magnitude expected from the Navier-Stokes form of Ohm's law. A large relaxation time leads to a faster decay of the magnetic field at early times, but the induced current persists longer, which can then reverse the decay and lead to an increase of the magnetic field at later times.
Furthermore, when the product of relaxation time and conductivity is large, the equation of motion for the charge diffusion current corresponds to that of an underdamped harmonic oscillator, which leads to an oscillatory behavior of the charge diffusion current. 
We also found that, in the case of an initially expanding fluid, the back-reaction of the magnetic fields onto the fluid crucially depends on the value of the conductivity and the magnitude of the inverse plasma $\beta$-parameter. For small $\beta^{-1} \ll 1$, the back-reaction is negligible, no matter whether we choose a large or a small conductivity. Only when $\beta^{-1} \gg 1$, we see a sizable back-reaction, which is larger for a smaller value of the conductivity.

There are many possible directions for future work. First, one could relax the assumption of transverse homogeneity and extend this to a full 3+1-dimensional situation to have a realistic comparison with experimental data. Second, one could include a finite initial net electric charge distribution and study the interplay between electromagnetic field and fluid in the context of net charge fluctuations and their dissipation. This is important for the Beam Energy Scan program at RHIC. Finally, one could extend the present treatment to allow the matter to have a finite polarization and magnetization. These and further questions will be addressed in future work.

\begin{acknowledgments}
We thank G.\ Inghirami and M.\ Mayer for fruitful discussions. A.D gratefully acknowledges the support of the Alexander von Humboldt Foundation through a research fellowship for postdoctoral researchers.
This work is supported by the Deutsche Forschungsgemeinschaft
(DFG, German Research Foundation) through the Collaborative Research
Center CRC-TR 211 ``Strong-interaction matter under extreme conditions''
- project number 315477589 - TRR 211 and by the State of Hesse within the Research Cluster
ELEMENTS (Project ID 500/10.006).
\end{acknowledgments}

\appendix

\section{IMEX scheme for relativistic resistive second-order dissipative MHD}
\label{sec:IMEX}

In what follows we provide a more detailed discussion of our numerical implementation of the IMEX method for relativistic resistive second-order dissipative MHD. We start by recalling that, in general, an IMEX scheme applies an implicit discretization scheme to the stiff terms and an explicit one to the non-stiff ones. For the system \eqref{eq:prototype}, it takes the form \cite{Pareschi:2005:IER,2013rehy.book.....R}
\begin{eqnarray}
 \bar{\boldsymbol{U}}^{i}&=&\boldsymbol{U}^{n}-\Delta t\sum_{k=1}^{i-1}\tilde{a}_{ik}\left[
 \partial_j \boldsymbol{F}^j(\bar{\boldsymbol{U}}^{k})- \boldsymbol{T}(\bar{\boldsymbol{U}}^k) \right]+\Delta t\sum_{k=1}^i a_{ik} \frac{1}{\epsilon}\boldsymbol{R}(\bar{\boldsymbol{U}}^{k}) \;,\\
 \boldsymbol{U}^{n+1}&=&\boldsymbol{U}^n-\Delta t\sum_{i=1}^{\nu}\tilde{\omega}_i \left[\partial_j \boldsymbol{F}^j(\bar{\boldsymbol{U}}^{i})-\boldsymbol{T}(\bar{\boldsymbol{U}}^i)\right]+\Delta t\sum_{i=1}^\nu \omega_{i}\frac{1}{\epsilon}\boldsymbol{R}(\bar{\boldsymbol{U}}^{i}) \;,
\end{eqnarray}
where $\bar{\boldsymbol{U}}^{i}$ is the vector of conserved quantities at the intermediate time steps $i$ of the RK scheme, $i = 1, \ldots , \nu$, while $\boldsymbol{U}^n$ is the vector of conserved quantities at time step $t^n$, where $n$ is the number of time steps. The matrices $\tilde{A} = (\tilde{a}_{ik} )$ and $A = (a_{ik} )$ are $\nu \times \nu$ matrices such that the resulting scheme is explicit in $\partial_j \boldsymbol{F}^j- \boldsymbol{T}$ (i.e., $\tilde{a}_{ik} = 0$ for $k \geq i$) and implicit in $\boldsymbol{R}$
(i.e., $a_{ik} = 0$ for $k > i$). A specific IMEX
scheme is characterized by these two matrices and the coefficient vectors $\tilde{\omega}_i$ and $\omega_i$.

A convenient way of describing an IMEX scheme is offered by the Butcher notation, in which the scheme is given by two  tableaux (one for the explicit and the other for the implicit time-stepping) of the type \cite{2013rehy.book.....R}
\begin{align}
&
\renewcommand\arraystretch{1.2}
\begin{array}
{c|c}
\tilde{c} & \tilde{A}\\
\hline
& \tilde{\omega}^T\\
\end{array}&
\renewcommand\arraystretch{1.2}
\begin{array}
{c|c}
{c} & {A}\\
\hline
& {\omega}^T\\
\end{array}
\end{align}
where the index $T$ indicates transposition and where the coefficient
vectors $\tilde{c}$ and $c$ satisfy the constraints
\begin{align}
    \tilde{c}_i&=\sum^{i-1}_{j=1}\tilde{a}_{ij}\;, &c_i&=\sum^{i}_{j=1}a_{ij}\;.
\end{align}

A viable numerical scheme maintains so-called strong stability at the discrete level and is then called Strong Stability Preserving (SSP), see Ref.\ \cite{10.2307/4100965} for a detailed description of optimal SSP schemes and their properties. In the present work we use the SSP3 (3,3,2) scheme for our numerical solution \footnote{In this context, the abbreviation $\mathrm{SSP}k $ $(s,\sigma,p)$ means the following. The index $k$ denotes the order of the SSP scheme. The triplet of numbers $(s,\sigma,p)$ indicates with $s$ the number of stages of the implicit scheme (in our case $s = \nu$), with $\sigma$ the number of stages of the explicit scheme (in our case $\sigma = \nu$), and with $p$ the order of the IMEX scheme.}. The Butcher-tableau form  corresponding to this scheme is adopted from Refs.\ \cite{Pareschi:2005:IER,2013rehy.book.....R} and given in Table \ref{table:TableIMEX}. 

For the system \eqref{Eq:systemofeqn} of equations of motion it is possible to introduce a natural decomposition of variables in terms of stiff and non-stiff parts. The vector $\boldsymbol{U}$ can be split into two subsets: the stiff terms $\boldsymbol{X}=(V^i,E^i)^T$ and the non-stiff terms $\boldsymbol{Y}=(e,M^i,N_f,B^i)^T$. This choice is natural because the system has two intrinsic timescales $\sigma$ and $\tau_V$. When $\tau_V\rightarrow 0$, we recover the usual Navier-Stokes form of Ohm's law and the fields decay according to the timescale set by the conductivity $\sigma$, whereas when $\tau_V$ is finite the evolution of the system is dictated by the interplay of these two timescales and can be characterized by the value of the damping coefficient $\zeta_d$ entering Eq.\ \eqref{eq:DampedOscii}.
\begin{table}
\caption{Butcher tableaux for the explicit and implicit SSP3 (3,3,2) scheme.}
\medskip 
\maketabularEx
\hspace{0.1\textwidth}
\maketabularIm
\label{table:TableIMEX}
\end{table}
As a result, the procedure to compute each intermediate value $\bar{\boldsymbol{U}}^{i}$ of the IMEX scheme can be performed in two steps:
\begin{enumerate}
 \item  Perform an explicit time step by computing intermediate values $\boldsymbol{Y}^i,\boldsymbol{X}^i$, where $i$ labels the RK step,
 \begin{eqnarray}
  \boldsymbol{Y}^i&=&\boldsymbol{Y}^n-\Delta t\sum_{k=1}^{i-1}\tilde{a}_{ik}\left[\partial_j \boldsymbol{F}^j_Y(\bar{\boldsymbol{U}}^{k})-\boldsymbol{T}_Y(\bar{\boldsymbol{U}}^k)\right] \;,\\
  \boldsymbol{X}^i&=&\boldsymbol{X}^n-\Delta t\sum_{k=1}^{i-1}\tilde{a}_{ik} \left[\partial_j \boldsymbol{F}^j_X(\bar{\boldsymbol{U}}^{k})-\boldsymbol{T}_X(\bar{\boldsymbol{U}}^k)\right] +  \Delta t\sum_{k=1}^{i-1}\frac{{a}_{ik}}{\epsilon^{k}}\boldsymbol{R}_X(\bar{\boldsymbol{U}}^{k})\;, \label{eq:A19}
 \end{eqnarray}
 where $ \boldsymbol{F}^j_X$, $\boldsymbol{F}^j_Y$
 are the fluxes pertaining to the stiff and non-stiff terms, respectively, and $\boldsymbol{T}_Y$, $\boldsymbol{T}_X$ are the non-stiff source terms corresponding to the variables $\boldsymbol{Y}$ and $\boldsymbol{X}$, respectively.
 $\boldsymbol{R}_X$ is the source term for the stiff variables, and the relaxation parameter $\epsilon^{k}\equiv\epsilon(\boldsymbol{Y}^k)$ is in principle also allowed to depend  on the non-stiff $\boldsymbol{Y}^k$ variables at the 
 $k^{\text{th}}$ RK step.
\item Perform an implicit time step, which involves only $\boldsymbol{X}$, by solving
\begin{eqnarray}
   \bar{\boldsymbol{Y}}^{i}&=&\boldsymbol{Y}^i\;,\\
\bar{\boldsymbol{X}}^{i}&=&\boldsymbol{X}^i+\Delta t\frac{a_{ii}}{\epsilon^{i}}\boldsymbol{R}_X(\bar{\boldsymbol{U}}^{i})\;.
\end{eqnarray}
\end{enumerate}
 The implicit equation for $\bar{\boldsymbol{X}}^i=(\bar{V}^i,\bar{E}^i)^T$  can be solved,
  yielding
 \begin{align}\label{eq:invertMa}
   \bar{V}^i&={\left(1+\frac{\Delta t\, a_{ii}}{\tau_V\gamma} \right)}^{-1}\left(V^{i}+\frac{\Delta t \, a_{ii}}{\tau_V\gamma}\frac{\sigma}{q} \bar{\mathcal{E}}^i\right)
  \end{align}
and
\begin{equation}\label{eq:invertMa2}
 \bar{E}^i=E^{i}-\Delta t \, a_{ii}\left[\bar{N}_f v^i +q\left(\delta^{ij} + v^i v^j \right) \bar{V}^j \right]\;,
\end{equation}
where we have used the identity $V^\mu u_\mu=0$. Equations (\ref{eq:invertMa}) and (\ref{eq:invertMa2}) have to be solved iteratively for $(\bar{V}^i, \bar{E}^i)$. It is interesting to note that the implicit solution \eqref{eq:invertMa} is consistent with the known solutions in the following two limits. In the limit $\tau_V\rightarrow 0$, the term
$V^i$ on the right-hand side can be neglected, yielding the Navier-Stokes form of Ohm's law, $q\bar{V}^i = \sigma \bar{\mathcal{E}}^i$. 
On the other hand, in the limit of $\tau_V\rightarrow \infty$, the implicit step is trivial, $\bar{V}^i = V^i$, and the charge diffusion current decouples from the electric field.

One crucial step remains before we arrive at the final solution of the system \eqref{Eq:systemofeqn} of equations of motion. We describe this in the following. We note that the solution of the conserved quantities $\boldsymbol{Y}=(e,M^i,N_f,B^i)$ at time $t = (n + 1)\Delta t$ is obtained by simply evolving Eqs.\ \eqref{Eq:systemofeqn} using the explicit time-stepping. However, after the explicit time step, the diffusion current and electric field only assume an approximate solution $(V^i,E^i)$ according to Eq.\ \eqref{eq:A19}. The implicit inversion  step, giving rise to Eqs.\ \eqref{eq:invertMa} and \eqref{eq:invertMa2}, depends on the velocity $v^i$, which is not explicitly known, even after the explicit time step. The way to determine $v^i$, as well as other variables in the local rest frame, such as the energy density and particle number density (which are needed to compute the pressure $P$ via the equation of state), is similar to the one presented in Refs.\ \cite{Palenzuela:2008sf, Dionysopoulou:2012zv}. It is essentially a nested fixed-point iteration procedure and can be summarized by the following steps:
\begin{enumerate}
    \item Adopt as an initial guess for the velocity its value at the previous time step, which we denote by $v_\star^i$. The diffusion current and electric field $(\bar{V}^{i},\bar{E}^{i})$ are then computed via Eqs.\ \eqref{eq:invertMa} and \eqref{eq:invertMa2} as functions of $(V^i,v_\star^i,E^i)$ via fixed-point iteration.
    \item Subtract the Poynting flux and the electromagnetic energy density from the conserved variables, and define new variables as follows:
    \begin{align}
        M^{\prime i}&=M^i - \epsilon^{ijk} E_j B_k \equiv \gamma^2 w v^i\;,\\
        e^{\prime}&=e - \frac{1}{2} (E^2 + B^2) \equiv \gamma^2 w-P\;.
    \end{align}
The variables $(e^{\prime }, M^{\prime i})$ simply correspond to the conserved variables in ideal relativistic fluid dynamics and the velocity $v^i$ is computed by the standard fixed-point iteration procedure often used in the literature, see, e.g., Refs.\ \cite{Molnar:2009tx,Palenzuela:2008sf}.
\item Replace $v_\star^i$ with the velocity $v^i$ obtained from step 2., and repeat steps 1.--3., until the  variables $(\bar{V}^i, v^i, \bar{E}^i)$ have converged within a given tolerance limit (in our case $10^{-7}$). 
\end{enumerate}
The approach discussed above is a simple procedure that can be implemented straightforwardly and works well if the inverse plasma $\beta$-parameter $B^2/(2P)$ is not too large. In the cases studied by us, convergence was reached within 10 to 50 iterations depending on the values of $\sigma$ and $\tau_V$. 

\section{Test cases}\label{App:TestCases}

In this section, two one-dimensional test cases (see, e.g., Refs.\  \cite{Komissarov:2007wk, Palenzuela:2008sf,Takamoto:2011ng}) are presented. For all results presented here, the fluxes $\boldsymbol{F}^j_{X,Y}$ are determined via the Godunov-type Harten–Lax–van Leer–Einfeldt (HLLE) algorithm \cite{doi:10.1137/0721001}. As in ideal MHD, the numerical calculation implements total-variation-diminishing (TVD) methods for the reconstruction of the solution.  We use an ideal equation of state $P =(\Gamma- 1)\varepsilon$, where $\Gamma$ is the adiabatic index. We use an extreme $\Gamma = 2$, which corresponds to the  speed of sound $c_s^2\equiv\Gamma-1=1$.
Our time step is $\Delta t = \lambda_{\text{CFL}} \Delta x$, with
Courant-Friedrichs-Lewy number $\lambda_{\text{CFL}} = 0.1$.

\subsection{Shock-tube problem}\label{sec:ShockTube}

  For the shock-tube problem with discontinuity at $x=0$ we take
  the initial pressure and magnetic field on the left ($L$)- and right ($R$)-handed side (in dimensionless units) as
 \begin{align}
  (P^L,B_y^L)&=(1.0,0.1)\;,\\
  (P^R,B_y^R)&=(0.1,-0.1)\;,
 \end{align}
 while other fields are set to 0. We compare our results to the exact solution of the ideal-MHD Riemann problem presented in Ref.\ \cite{Giacomazzo:2005jy}. The calculation runs from initial time $t=0$ to the final time $t=0.4$. The grid spacing is set to $\Delta x=0.002$.
   \begin{figure}
    \centering
    \includegraphics[width=0.45\textwidth]{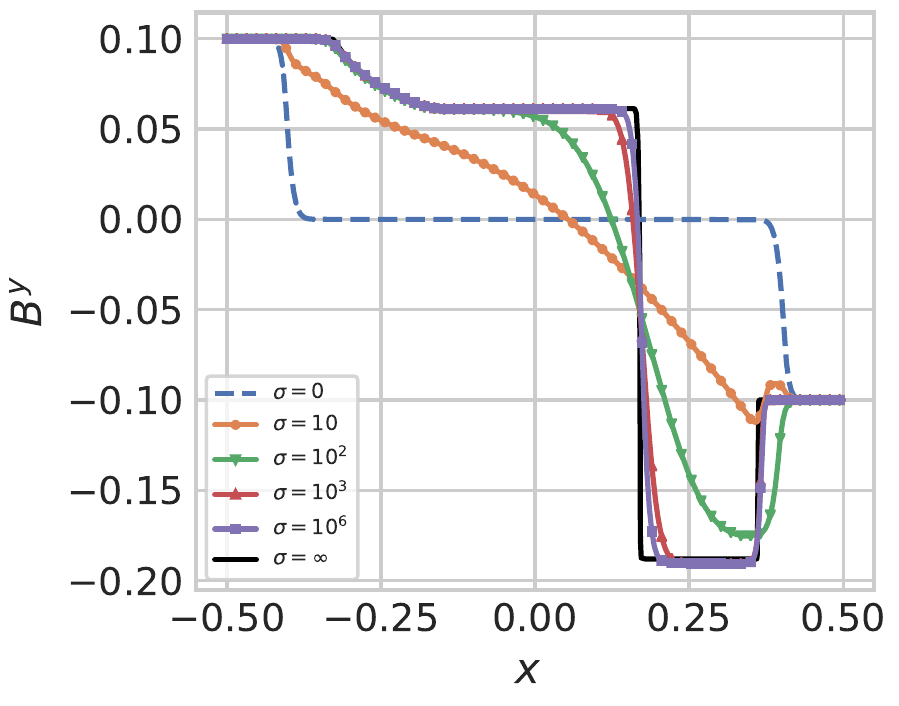}
    \includegraphics[width=0.45\textwidth]{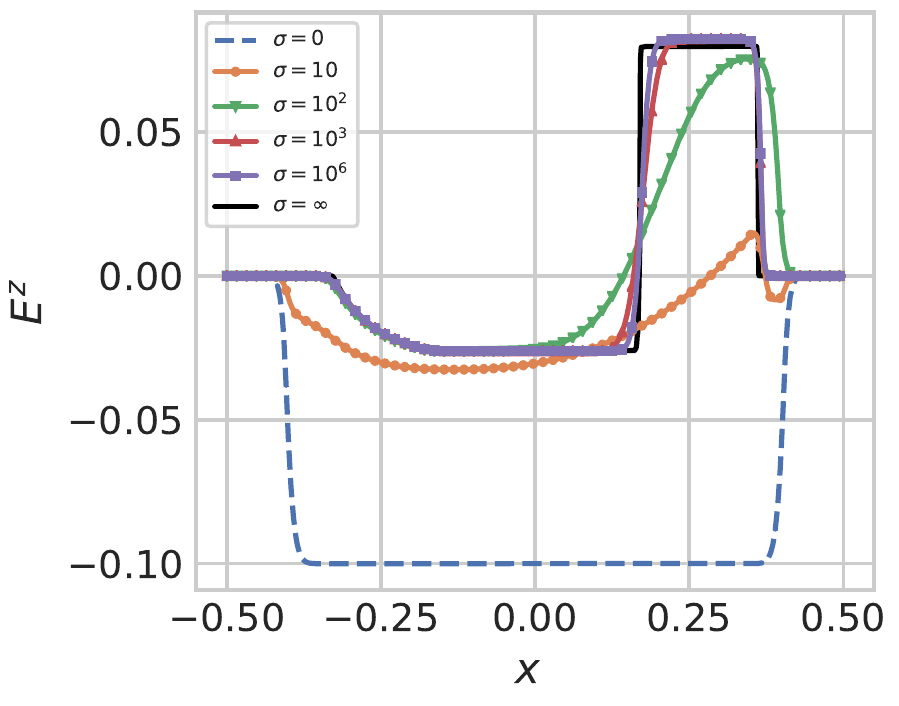}
    \includegraphics[width=0.45\textwidth]{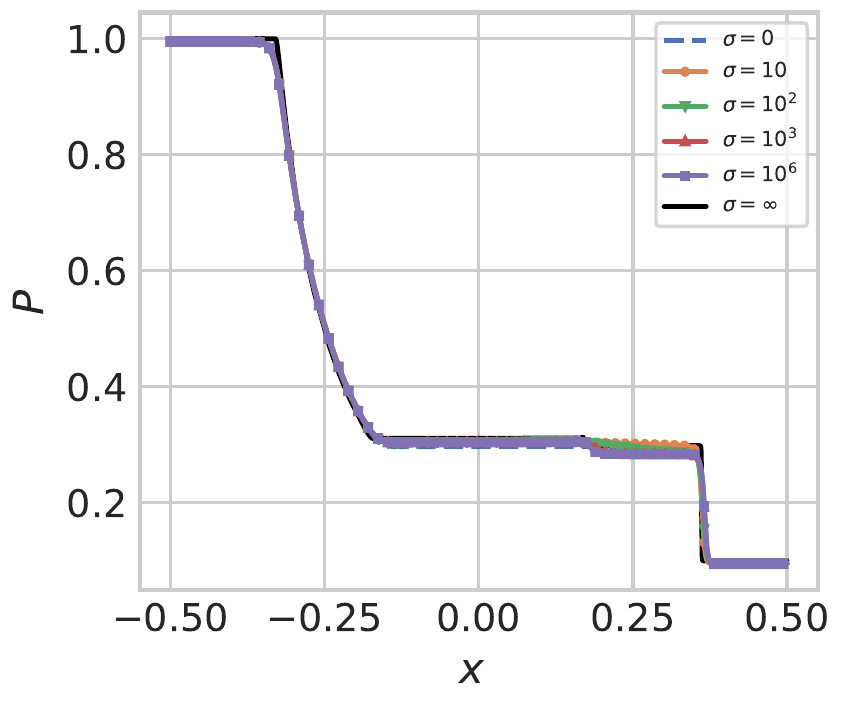}
    \includegraphics[width=0.45\textwidth]{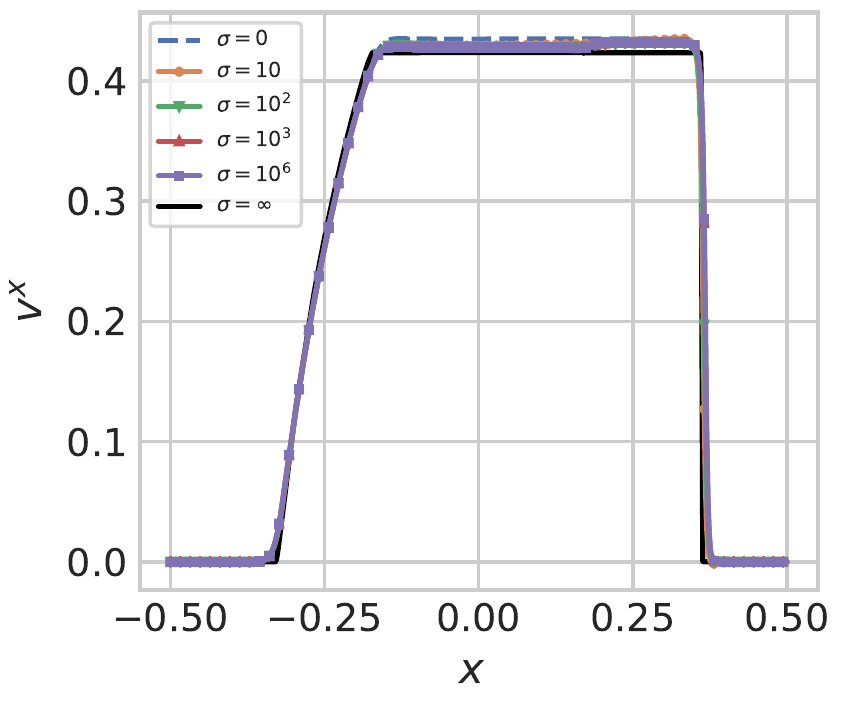}
    \caption{Top row:  Magnetic-field component $B^y$ (left panel) and electric-field component $E^z$ for the shock-tube problem at $t= 0.4$. Different lines refer to different values of the conductivity for $\tau_V = 10^{-3}$.  Bottom row: Same as top row, but for the pressure $P$ (left panel)  and  the fluid velocity $v^x$ (right panel).}
    \label{fig:Shocktubetest1}
\end{figure}
   \begin{figure}
    \centering
    \includegraphics[width=0.45\textwidth]{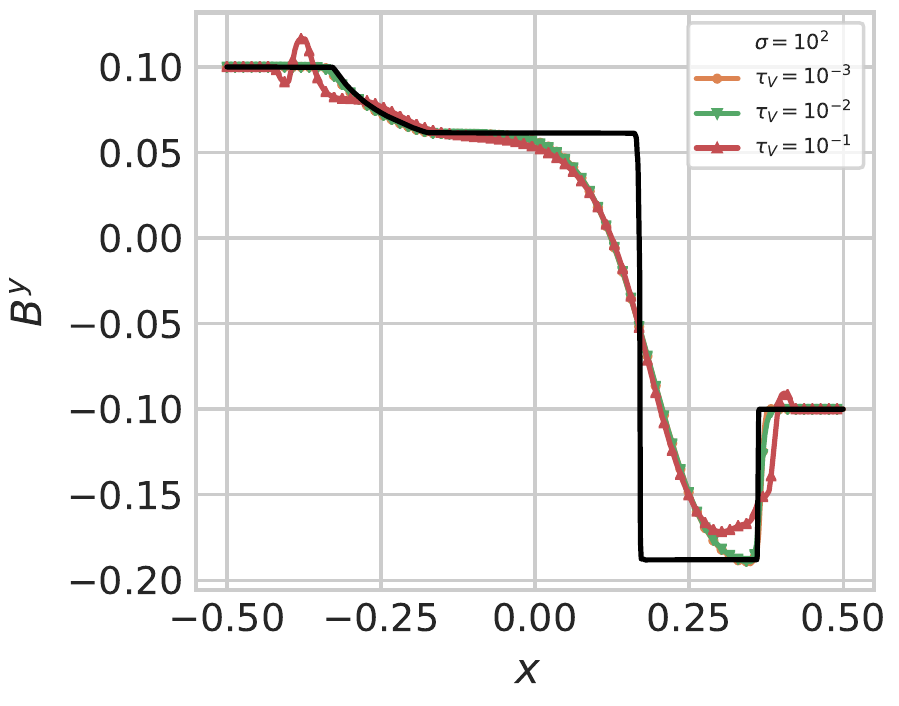}
    \includegraphics[width=0.45\textwidth]{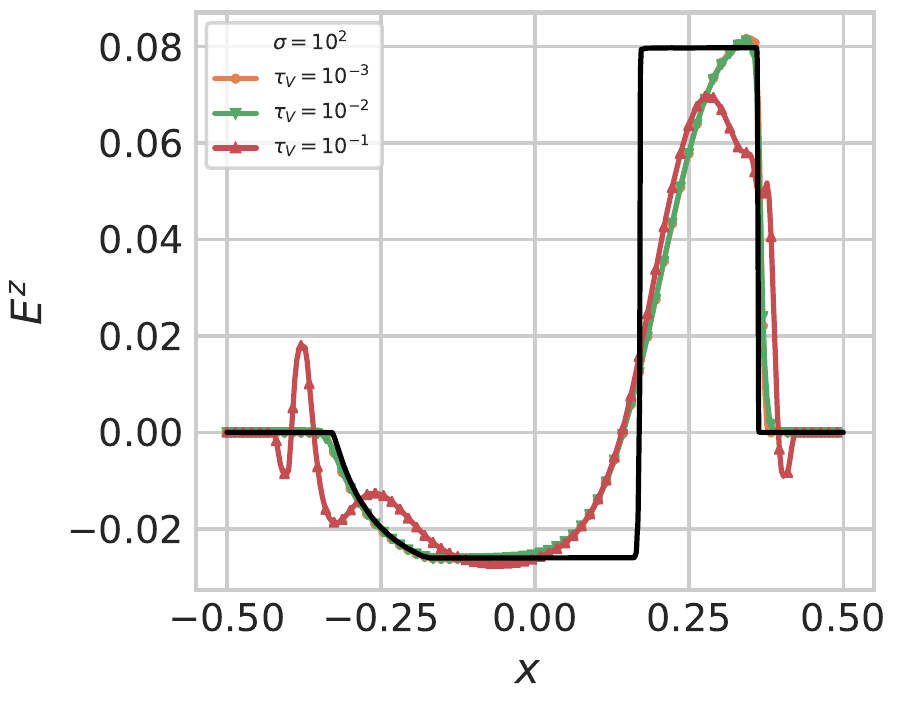}
    \caption{Same as Fig.\ \ref{fig:Shocktubetest1}, but now with constant $\sigma=10^2$ and different $\tau_V$.}
    \label{fig:Shocktubetest2}
\end{figure}

In Fig.\ \ref{fig:Shocktubetest1}, we take the relaxation time $\tau_V=10^{-3}$ to be small in order to compare the results with the Navier-Stokes form of Ohm's law. The electrical conductivity ranges from $\sigma=0$ corresponding to vacuum to $\sigma=10^6$, approximately corresponding to the ideal-MHD case. The top row of Fig.\ \ref{fig:Shocktubetest1} shows the magnetic field (left panel) and the electric field (right panel) at time $t=0.4$. One notices that for $\sigma = 0$ the solution describes a discontinuity propagating at the speed of light to the left and the right, corresponding to the solution of Maxwell's equations in vacuum. As the conductivity increases, the solution approaches the ideal-MHD one. 
The bottom row of Fig.\ \ref{fig:Shocktubetest1} shows the pressure (left panel) and the velocity (right panel). Since the inverse plasma $\beta$-parameter for this set-up is small, the effect of the magnetic field on these fluid-dynamical variables is small, and all curves resemble those from the ideal-MHD Riemann problem.

In Fig.\ \ref{fig:Shocktubetest2} the initial conditions are the same as for Fig.\ \ref{fig:Shocktubetest1}, except that we now solve the Riemann problem for a constant conductivity $\sigma=10^2$ and different relaxation times $\tau_V$. The two cases of small  $\tau_V = 10^{-2}$ and $10^{-3}$ yield almost identical results,  but for larger $\tau_V$ we see  oscillations appearing in the solution, which can be traced back to the fact that in this regime the diffusion current is underdamped, which then back-reacts to the electromagnetic fields.
   \begin{figure}
    \centering
    \includegraphics[width=0.45\textwidth]{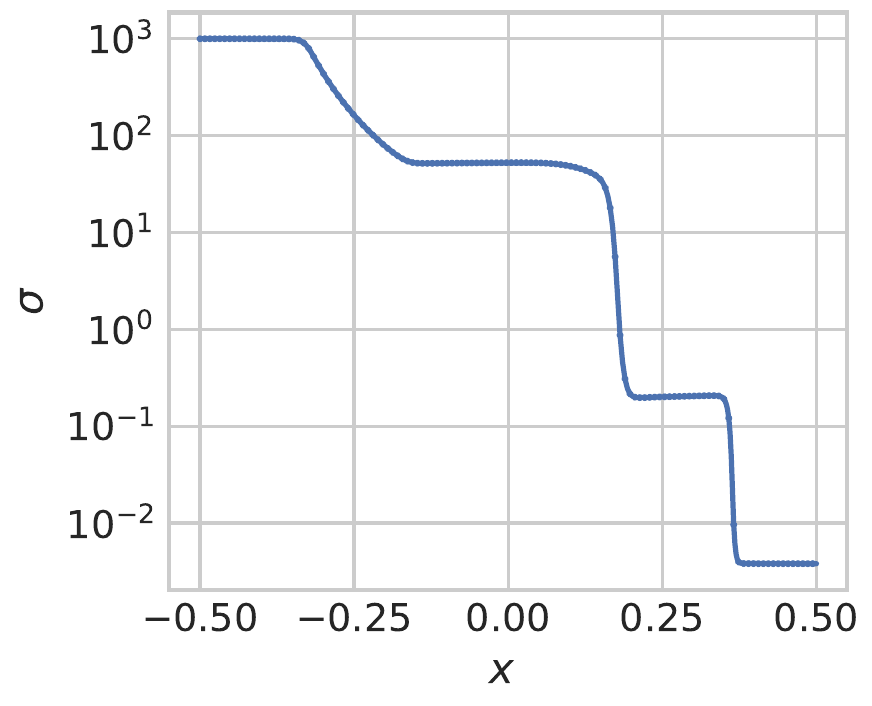}
    \includegraphics[width=0.45\textwidth]{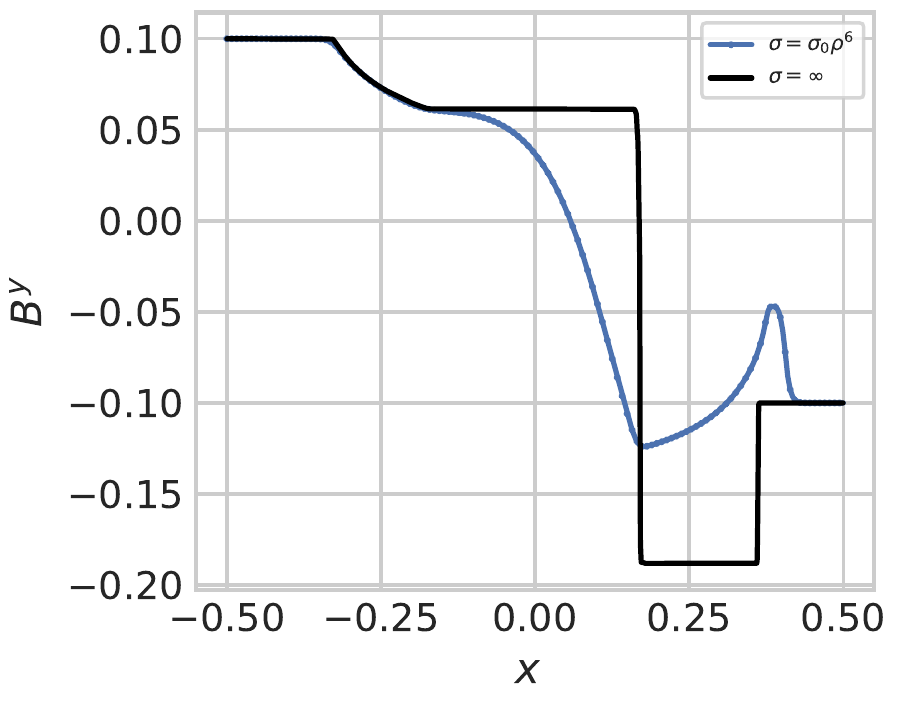}
    \caption{Left panel: Profile of a non-uniform conductivity $\sigma$ for the shock-tube problem. Right panel: The magnetic-field component $B^y$ for the conductivity profile as in the left panel. }
    \label{fig:Shocktubetest3}
\end{figure}
    \begin{figure}
    \centering
    \includegraphics[width=0.45\textwidth]{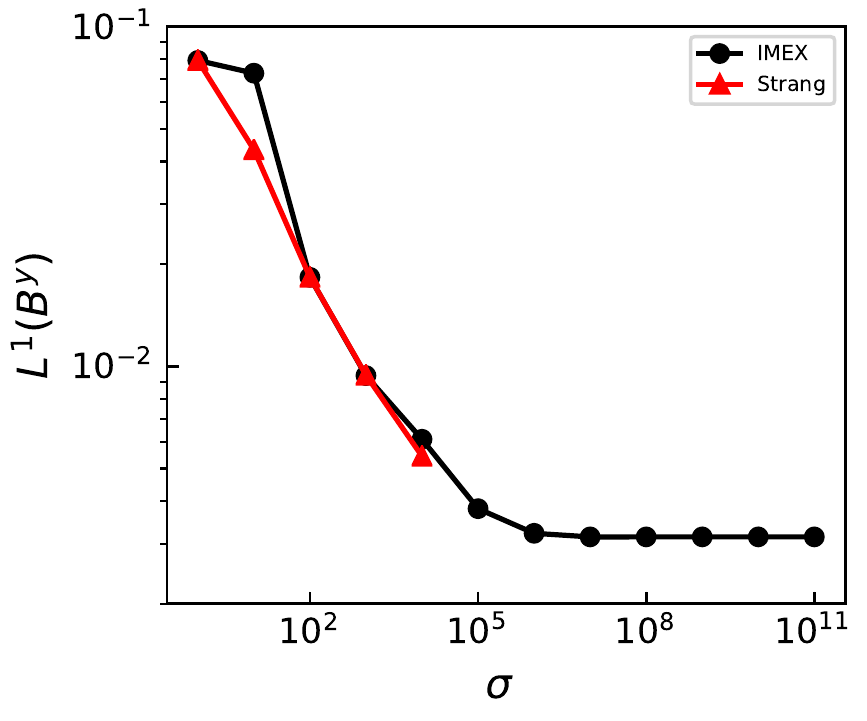}
    \caption{$L^1(B^y)$ norm of the magnetic field component $B^y$ between the
    numerical solution computed from the Strang-splitting technique and the IMEX schemes and the exact solution of the shock-tube in the ideal-MHD limit as a function of conductivities. The Strang-splitting technique does not yield a stable solution for conductivities values larger than $10^4$.}
    \label{fig:L1Norm}
\end{figure}
    \begin{figure}
    \centering
    \includegraphics[width=0.45\textwidth]{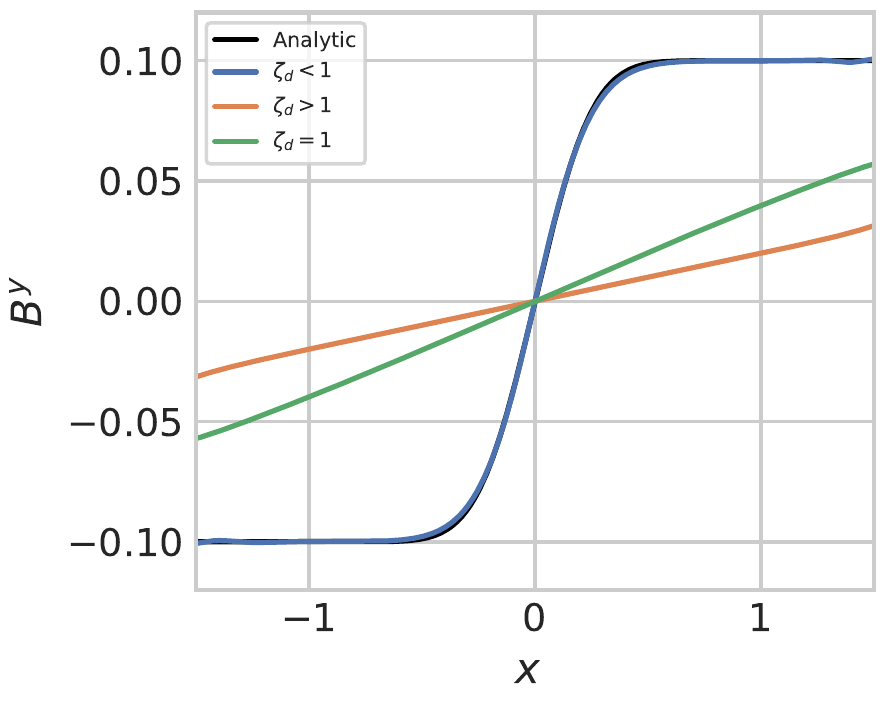}
    \includegraphics[width=0.45\textwidth]{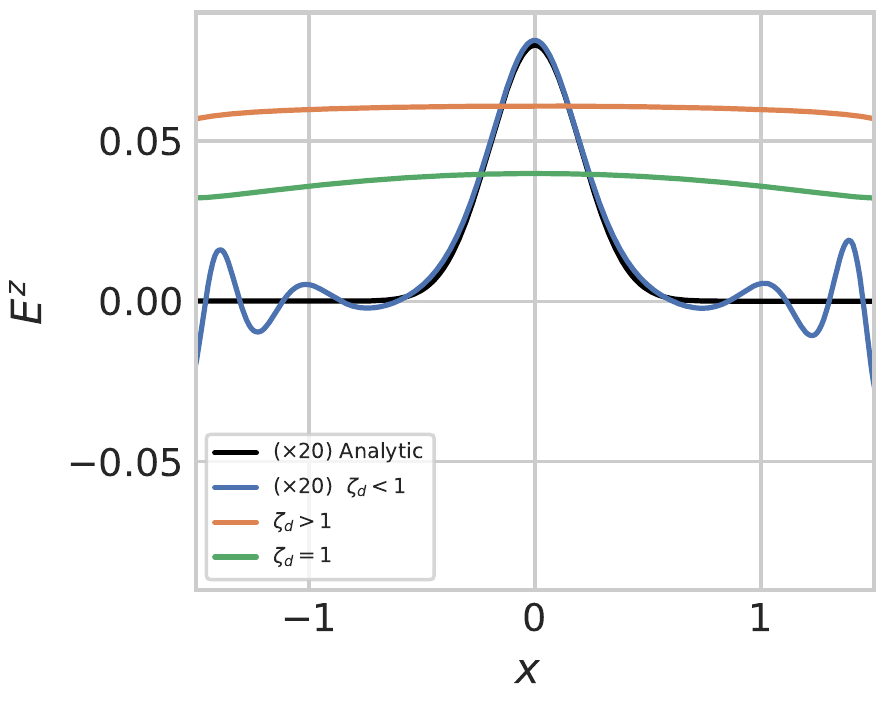}
    \caption{Left panel: The magnetic-field component $B^y$  at $t=2$ for the test case of a self-similar current sheet, for three different values of the damping ratio $\zeta_d = 1/(2 \sqrt{\sigma \tau_V})$. For the underdamped case ($\zeta_d <1$) we took $\sigma = 100$, for the overdamped case ($\zeta_d > 1$) we took $\sigma = 0.5$, while for the critical case
    ($\zeta_d =1$) we have $\sigma = 1.25$. For all cases $\tau_V = 0.2$. The black solid line is the analytic solution discussed in the text. Right panel: Same as in the left panel, but for the electric-field component $E^z$. }
    \label{fig:Currentsheet1}
\end{figure}

Further, in Fig.\ \ref{fig:Shocktubetest3}
we consider the same initial condition as in the previous two cases, but now with a  non-uniform conductivity, $\sigma=\sigma_0 \rho^\delta$, where
$\rho$ is a solution of the
advection equation $\partial_t \rho+
\partial_j (\rho v^j) =0$ with initial
condition $\rho^L = 1.0$, $\rho^R = 0.125$. Furthermore, $\sigma_0=10^3$, $\tau_V=10^{-3}$, and $\delta=6$. The resulting conductivity $\sigma$ and $B^y$ at time $t=0.4$ are shown in Fig.\ \ref{fig:Shocktubetest3}. 
It should be stressed
that because of the functional dependence of $\sigma$ on $\rho$, the region on the left, i.e., $x<0$, has a very high conductivity at this time and, thus, the numerical solution tends to the ideal-MHD case. The opposite happens in the right region, i.e., $x>0$, where the conductivity is lower and the solution tends towards the vacuum one. The results presented in Fig.\ \ref{fig:Shocktubetest3} show that our algorithm can handle non-uniform conductivity profiles even in the presence of shocks.

Finally, we also perform a comparison between the IMEX and the Strang-splitting approaches. In Fig.\ \ref{fig:L1Norm} we show the $L^1$-norm of the difference between the numerical solution obtained from both schemes with the ideal-MHD exact solution, for different values of the conductivity with $\Delta x=0.003$ and $\tau_V=10^{-3}$ (in dimensionless units). The $L^1(B^y)$ is defined as
\begin{equation}
    L^1(B^y)=\frac{1}{N}\sum_i^N|B^y(x_i)-B_{\mathrm{id-MHD}}^y(x_i)|\;,
\end{equation}
 where $N$ is the total number of cells. Firstly, the reported difference between the numerical solution for the resistive MHD equations
and the ideal-MHD equations should not be interpreted as an error given that the ideal-MHD solution is not the correct solution of the equations at finite conductivity. This is particularly prominent for the lower conductivity values and improves as we increase the value of conductivity. Furthermore, we did not find any numerical instability in the IMEX method for $\sigma$ ranging from $0$ to $10^{11}$ within the above resolution. For the Strang-splitting technique the solutions becomes unstable and no numerical solution could be obtained already for moderately high values of the conductivity, i.e, beyond $\sigma>10^4$. Our results are in agreement with similar conclusions found previously in Ref.\ \cite{Palenzuela:2008sf}. Also, we note that the difference between the numerical solution for the IMEX scheme and the exact ideal-MHD solution saturates between $\sigma\sim 10^5-10^6$. This emphasizes the fact that to gain more accuracy one needs to increase the resolution.

\subsection{Self-similar current sheet}

 In this test case, it is assumed that the magnetic pressure is much smaller than that of the fluid, i.e., $B^2\ll P$, so that the background fluid is not influenced by the evolution of the magnetic field. We assume that the magnetic field has the form $\boldsymbol{B} = (0, B^y(x,t), 0)$, with $B^y(x,0) = B_0\
 \textrm{sgn}(x)$, while the fluid pressure is constant. If one assumes that the conductivity $\sigma$ is large, one can find an approximate solution for the magnetic field \cite{Komissarov:2007wk,Palenzuela:2008sf}
 \begin{equation}\label{eq:SelfsimB}
  B^y(x,t)=B_0\,\mathrm{erf}\left(\frac{1}{2}\sqrt{\frac{\sigma}{\xi}}\right)\;,
 \end{equation}
 while the electric field evolves as
  \begin{equation}\label{eq:SelfsimE}
  E^z(x,t)=\frac{B_0}{\sqrt{\pi\sigma t}}\, \mathrm{exp}\left(-\frac{\sigma}{4\xi}\right)\;,
 \end{equation}
 where $\xi=t/x^2$. Note that in the derivation of the above solution one has neglected the displacement current (thus the evolution equation of the magnetic field is not hyperbolic any more) and one has used the Navier-Stokes form of Ohm's law. Although only an approximate solution, comparing it with the numerical solution will allow us to discuss various regimes of our resistive MHD description.

 We initialize the electromagnetic fields according to Eqs.\ \eqref{eq:SelfsimB} and \eqref{eq:SelfsimE} at time $t=10^{-3}$ with $B_0=0.1$ and constant fluid pressure $P=40.0$, along a spatial grid ranging from $[-3.5,3.5]$ with $\Delta x=0.001$. The left panel of Fig.\ \ref{fig:Currentsheet1} shows the magnetic field at time $t=2$ for three values of the damping ratio $\zeta_d$. Since the solution \eqref{eq:SelfsimB} is valid for large values of $\sigma$ the analytical and numerical results match very well for $\zeta_d<1$. For $\zeta_d>1$ and $\zeta_d=1$, the numerical solution is more diffusive and thus not captured by the analytical solution. The right panel of Fig.\ \ref{fig:Currentsheet1} shows the electric field. In the underdamped case $\zeta_d<1$ we observe oscillations at the left and right edges of the grid. These oscillation die out at late times and the analytical and numerical results are in perfect agreement.
 
\section{Equations of motion in Milne coordinates}\label{App:Milnecord}

In Milne coordinates, and assuming transverse homogeneity, the system  of equations of motion (\ref{eq:partialTmunu}), (\ref{eq:partialrho}), (\ref{eq:maxwelleqn1}), (\ref{eq:maxwelleqn2}), and (\ref{eq:Ohmslaw}) of relativistic resistive second-order dissipative MHD can be cast into the following form 
\begin{equation}\label{Eq:systemofeqnMilne}
 \partial_\tau(\boldsymbol{U})+\partial_\eta(\boldsymbol{F}^\eta(\boldsymbol{U}))=\boldsymbol{{S}}(\boldsymbol{U})\;,
\end{equation} 
where $\boldsymbol{U}$ represents the vector of conserved variables and $\boldsymbol{F}^\eta$ that of the fluxes, 
$$\boldsymbol{U}=\begin{pmatrix}
 \tau e \\ \tau M^x\\\tau M^y\\ \tau^3 M^\eta \\N_f\\ V^i \\\tau\tilde{B}^y\\\tau \tilde{B}^x \\\tau \tilde{E}^y\\\tau \tilde{E}^x
\end{pmatrix} \;, \qquad \boldsymbol{F}^\eta(\boldsymbol{U})=
\begin{pmatrix}
 \tau F^\eta_e\\ \tau F^{x\eta}_M \\ \tau F^{y\eta}_M \\\tau^3 F^{\eta\eta}_M  \\ v^\eta N_f\\v^\eta V^i\\ \tilde{E}_x\\ -\tilde{E}_y\\-\tilde{B}_x\\\tilde{B}_y
\end{pmatrix}\;,
$$ 
while the sources read
$$\boldsymbol{S}(\boldsymbol{U})=
\begin{pmatrix}
0\\0\\0\\-\tau^2  F^{\eta\eta}_M\\ -N_f/\tau -\partial_i(-v^iV^0+V^i)\\ 
\left(\sigma \mathcal{\tilde{E}}^i+q\kappa\nabla^{i}\alpha-qV^{i}\right)/(q \tau_V \gamma)+V^i\partial_j v^j-u^i V^\nu \dot{u}_\nu-G^i_n/\gamma\\0\\0\\-\tau J_f^y\\-\tau J_f^x 
\end{pmatrix}\;.
$$
Here, $G_n^i=u^\alpha\Gamma^i_{\alpha\beta}V^\beta$ are additional geometrical source terms, where $\Gamma^\mu_{\alpha\beta}$ are the Christoffel symbols for Milne coordinates. Additionally, we have the constraint equations
\begin{align}\label{eq:constEMilne}
\partial_\eta{\tilde{E}^\eta}&=0\;,\\\label{eq:constBMilne}
 \partial_\eta\tilde{B^\eta}&=0\;, 
\end{align}
where we have assumed that the system has zero net charge. The definition of the conserved quantities $e$, $M^i$, and $N_f$ and the fluxes $F^i_e$ and $F^{ij}_M$ are the same as in Cartesian coordinates. The definition of the electromagnetic fields with a tilde comes from the coordinate transformation \eqref{eq:CoordTransform} from Cartesian to Milne coordinates.
\bibliography{main}
\end{document}